\documentclass[useAMS,usenatbib]{mn2e}
\usepackage[pdftex,pdfpagemode={UseOutlines},bookmarks,bookmarksopen,colorlinks,linkcolor={blue},citecolor={blue},urlcolor={red}]{hyperref}
\usepackage{graphicx}
\usepackage{natbib}
\usepackage{xspace}
\usepackage{amsmath}
\usepackage{epsf}
\usepackage{widetext}
\usepackage{flushend}
\usepackage{cuted}
\usepackage{subfigure}
\usepackage{amssymb}
\usepackage{epstopdf}

\title[RSD from group-galaxy correlations]
{Group-galaxy correlations in redshift space as a probe of the growth of structure}

\author[F. G. Mohammad et. al.]{F. G. Mohammad\thanks{E-mail: faizan.mohammad@brera.inaf.it}$^{1,2}$, S. de la Torre$^3$, D. Bianchi$^{4,1}$, L. Guzzo$^{1}$, J. A. Peacock$^{5}$ \\
$^{1}$INAF-Osservatorio astronomico di Brera, via Emilio Bianchi 46, I-23807 Merate (LC), Italy \\
$^{2}$Universit\`a dell'Insubria, Dipartimento di Scienza e Alta Tecnologia, Via Valleggio 11, I�22100 Como, Italy\\
$^{3}$Aix-Marseille Universit\'e, CNRS, LAM (Laboratoire d'Astrophysique de Marseille) UMR 7326, 13388, Marseille, France \\
$^{4}$ICG - Institute of Cosmology and Gravitation, University of Portsmouth, Dennis Sciama Building, Burnaby Road, \\ Portsmouth, PO1 3FX, United Kingdom \\
$^{5}$Institute for Astronomy, University of Edinburgh, Royal Observatory, Blackford Hill, Edinburgh EH9 3HJ, United Kingdom}

\def\({\left(}
\def\){\right)}
\def\<{\left\langle}
\def\>{\right\rangle}
\def\xil{\xi^{s,(\ell)}(s)}
\def\xilt{\smash{\hat{\xi}^{s,(\ell)}(s)}}
\def\xirp{\xi^s(r_p,\pi)}
\def\mpcoh{\,h^{-1}{\rm Mpc}}
\def\[{\begin{equation}}
\def\]{\end{equation}}
\def\citejap#1{\citeauthor{#1}\ \citeyear{#1}}

\newcommand{\tablename}{Table} 
\newcommand{\tabref}[2][{}]{\hyperref[#2]{\tablename~\ref{#2}#1}} 

\newcommand {\x} {\! \times \!} 

\newcommand{\mnras}{{MNRAS}}
\newcommand{\apj}{{ApJ}}
\newcommand{\aj}{{AJ}}
\newcommand{\apjl}{{ApJL}}
\newcommand{\aap}{{A\&A}}

\newcommand{\prd}{{Phys.Rev.D}}

\begin{document}

\pagerange{\pageref{firstpage}--\pageref{lastpage}} \pubyear{2014}

\maketitle

\label{firstpage}

\begin{abstract} 
We investigate the use of the cross-correlation between galaxies and
galaxy groups to measure redshift-space distortions (RSD) and thus
probe the growth rate of cosmological structure.  This is compared to
the classical approach based on using galaxy auto-correlation. We make
use of realistic simulated galaxy catalogues that have been
constructed by populating simulated dark matter haloes with galaxies
through halo occupation prescriptions. We adapt the classical RSD
Dispersion model to the case of the group-galaxy cross-correlation
function and estimate the RSD parameter $\beta$ by fitting both the
full anisotropic correlation function $\xirp$ and its multipole
moments.  In addition, we define a modified version of the latter
statistics by truncating the multipole moments to exclude strongly
non-linear distortions at small transverse scales. We fit these three
observable quantities in our set of simulated galaxy catalogues and
estimate statistical and systematic errors on $\beta$ for the case of
galaxy-galaxy, group-group, and group-galaxy correlation functions.
When ignoring off-diagonal elements of the covariance matrix in the
fitting, the truncated multipole moments of the group-galaxy
cross-correlation function provide the most accurate estimate, with
systematic errors below 3\% when fitting transverse scales larger than
$10\mpcoh$. Including the full data covariance enlarges
  statistical errors but keep unchanged the level of systematic
  error.  Although statistical errors are generally larger for
groups, the use of group-galaxy cross-correlation can potentially
allow the reduction of systematics while using simple linear or
Dispersion models.

\end{abstract}


\section{Introduction}

Cosmological observations over the past 15 years have established that
the Universe is undergoing a late-time phase of accelerated expansion
\citep[e.g.][]{riess98,perlmutter99}. This is most
naturally explained with the presence of a `dark energy', a fluid with
negative pressure filling the entire Universe, which is currently
indistinguishable from Einstein's cosmological constant.
Alternatively, however, one may consider reproducing observations by
modifying the very nature of the gravitational equations of General
Relativity (e.g. \citejap{carroll04};
\citejap{joyce14}). Geometrical probes such as the Cosmic
Microwave Background (CMB), Baryon Acoustic Oscillations (BAO) and
type 1a Supernovae (SN1a) constrain the expansion history $H(z)$,
which however can be equally well fitted by both scenarios. This
degeneracy between dark energy and a modification of standard gravity
can only be lifted by dynamical probes looking at the growth of
structure inside the Universe, which is directly linked to the
underlying theory of gravity (\citejap{joyce14}).

The growth of cosmological structure induces galaxy peculiar
velocities, i.e. coherent flows of galaxies towards matter
overdensities.  When redshifts are used to map galaxy positions, we
are sensitive to such peculiar velocities, whose line-of-sight
component combines with the cosmological redshift. As a result, the
reconstructed spatial distribution of objects is distorted in the
radial direction, what are referred to as \emph{redshift-space
  distortions} (RSD). RSD can be quantified statistically in galaxy
redshift surveys by modelling the corresponding anisotropy that can be
measured in the two-point correlation function (2PCF)
$\xi(\mathbf{r})$ or, correspondingly, in the power spectrum
$P(\mathbf{k})$.  In the linear regime, this effect is described in
Fourier space by the Kaiser model \citep{kaiser87}, or its
equivalent in configuration space developed by
\citet{hamilton92}. Such modelling is however complicated by
the non-linear evolution of matter overdensities and velocities
\citep[see e.g.][and references therein]{delatorre12}.

On large scales we can still observe the linear, coherent flows
tracing the growth of cosmological structure, which enhance the
amplitude of two-point correlations.  But small-scale motions are
dominated by high-velocity galaxies in virialized structures,
resulting from the dynamical evolution of the highest peaks in the
density field.
The resulting stretching effect in galaxy maps is commonly referred to
as the \textit{Fingers of God} effect (FoG), due to the elongated
shapes that groups and clusters acquire when observed in redshift
space.
These two regimes of RSD produce characteristic features in the
two-point statistics of the galaxy distribution, which can be studied
by computing the correlation function as a function of two variables,
$r_p$ and $\pi$, perpendicular and parallel to the line of sight,
respectively. The two effects, linear and non-linear, introduce
respectively a large-scale squashing and a small-scale elongation
along the line-of-sight direction $\pi$. In linear theory, the
amplitude of the squashing effect is directly proportional to the
logarithmic growth rate of density fluctuations $f(z)$ (see
Eq. \eqref{eq:f}) \citep{kaiser87}. In practice, our discrete tracers will
generally be biased with respect to the overall matter distribution.
Under the hypothesis of a linear bias $b_{\rm lin}(z)$, the
large-scale squashing effect will depend on the parameter
$\beta(z)\equiv f(z)/b_{\rm lin}(z)$.

An empirical correction to the linear model was introduced as to
account for the contribution of nonlinear distortions, what is
normally referred to as the \emph{Dispersion model}
\citep{fisher94,peacock94}. In this model, the
linear expression is convolved (in configuration space), or multiplied
(in Fourier space) with a given distribution function for the pairwise
velocities $\varphi(v_{||})$ along the line of sight.  The dispersion
model has been widely and successfully used in the past to estimate
$\beta(z)$ from measurements of $\xi(r_p,\pi)$
\citep{peacock01, hawkins03, guzzo08,
  ross07, gaztanaga09a, gaztanaga09b,
  contreras13} or the power spectrum
\citep{percival04,
  tegmark04,blake11}. It has become clear that
such empirical correction provides estimates that can be biased by up
to 10\%, typically depending on the bias of the tracer being employed
\citep{okumura11,bianchi12}. The precision
  already achieved by current largest surveys requires to improve on
  these limitations to reach systematic errors of the order of the
  percent \citep[e.g. BOSS,][]{reid14, beutler14, samushia14, sanchez14, alam15}.  Such
limitations are intrinsic in the empirical nature of the dispersion
model \citep{scoccimarro04, delatorre12} and significant
effort has been invested over the past five years to improve this
theoretical description (e.g. \citejap{scoccimarro04};
\citejap{taruya10}; \citejap{reid11};
\citejap{delatorre12}, and references therein,
\citejap{reid14}). Alternatively, these
  theoretical limitations may be evaded if we are able to (a) identify
  galaxy tracers that are less affected by non-linearities, and/or (b)
  build novel statistics of galaxy clustering for which model fits are
  less sensitive to the same effect, while keeping the theoretical
  model as simple as possible.

In this paper we investigate two options along these two
avenues. Specifically, we consider the case in which one has access to
a robust galaxy group catalogue within the volume of a corresponding
galaxy redshift survey; this will be possible with next-generation
galaxy redshift surveys such as Euclid \citep{laureijs11},
also in combination with observations in other bands, as in particular
the X-ray data from e-Rosita (\citejap{merloni12}).  We
therefore first consider the merits of using the galaxy group-galaxy
cross-correlation to extract the distortion parameter $\beta$. We
might expect the centre of mass of a galaxy group to have little or no
random velocity, since the internal orbital velocities produce
Fingers-of-God only in the overall galaxy distribution.  In this view,
group-group correlations should be ideally placed (as also suggested
by the dependence of the systematic error on the halo mass shown by
\citejap{bianchi12}).  Moreover, the group-galaxy
cross-correlation provides a higher statistical power than using group
auto-correlation, given the higher number density of the galaxy
catalogue. Secondly, we investigate a new way of compressing
  the cosmological information present in the anisotropic two-point
  correlation function in redshift space, by building modified
  (truncated) multipole moments of $\xi(r_p,\pi)$ that partially
  mitigate small-scale non-linearities. These functions avoid
  explicitly including the contributions from small transverse scales,
  dominated by non-linear distortions\footnote{In a parallel work,
    \citejap{reid14} independently define a similar
    statistics as to mitigate fibre collisions effects in the analysis
    of RSD in the BOSS survey.}. In this analysis, we therefore
  compare the standard approach of using galaxy or group
  auto-correlation and the novel approach of using the galaxy-group
  cross-correlation, as well as study how the different estimators of
  the two-point statistics, namely the anisotropic 2PCF $\xirp$, its
  standard multipoles $\xi^{s,(\ell)}(s)$ and its truncated multipoles
  $\hat{\xi}^{s,(\ell)}(s)$, behave in recovering the distortion
  parameter $\beta$ in simulations.

The paper is organized as follows. In Section \ref{sec:model}, we derive
the linear Kaiser/Hamilton and Dispersion models for the
cross-correlation in the case of different observables: the anisotropic
2PCF $\xirp$, its multipole moments $\xil$, and the related truncated
multipole moments $\xilt$. In section \ref{sec:methodology}, we describe
the data, two-point correlation function estimators, as well as
ingredients needed to build theoretical models. In Section
\ref{sec:results_diag} we present the comparative analysis of the
different approaches and methods. Finally, in Section
\ref{sec:conclusion} we discuss our results and conclude.

\section{Redshift-Space Distortions: Modelling}	\label{sec:model}	

The apparent positions of objects are modified if we use redshifts to infer cosmological distances. The line-of-sight component of the peculiar velocity distorts positions in the following way:
	\begin{equation}
		\mathbf{s}=\mathbf{r}+\mathbf{u}(\mathbf{r})\cdot \hat{\mathbf{e}}_{||},      \label{eq:rtoz}
	\end{equation}
where $\mathbf{s}$ and $\mathbf{r}$ are objects positions, respectively, in redshift- and real space; $\hat{\mathbf{e}}_{||}$ is the unit vector along the line of sight; $\mathbf{u}$ is the scaled velocity field defined as $\mathbf{u}(\mathbf{r})=\mathbf{v}(\mathbf{r})/aH(a)$ with $\mathbf{v}(\mathbf{r})$ being the peculiar velocity field; $a$ is the scale factor; and $H(a)$ is the expansion rate of the Universe. The $\mathbf{s}$ coordinates constitute the so-called \textit{redshift space} and the distortions produced in the matter distribution are usually referred to as \textit{Redshift-Space Distortions} (RSD).

\subsection{Linear Model for the Cross-Correlation}\label{sec:kaiser}
	
We first derive the linear model for the two-point cross-correlation function. Following the derivation of \cite{kaiser87} for the auto-correlation, we start with assuming mass conservation in real- and redshift space in terms of overdensities $\delta(\mathbf{r})=\rho(\mathbf{r})/\langle{\rho}\rangle-1$:
		\begin{equation}
			\left[1+\delta_m^{s}(\mathbf{s})\right]d^3\mathbf{s}=\left[1+\delta_m(\mathbf{r})\right]d^3\mathbf{r},		\label{eq:masscon}
		\end{equation}
where $\delta_m(\mathbf{r})$ and $\delta_m^{s}(\mathbf{s})$ are, respectively, the matter density contrast in real- and redshift space. The volume element in redshift space $d^3\mathbf{s}$ is related to that in real space $d^3\mathbf{r}$ through $d^3\mathbf{s}=|J|d^3\mathbf{r}$ where $|J|$ is the Jacobian of the related coordinates transformation. In the particular case of the transformation in equation \eqref{eq:rtoz}, the Jacobian is given by
	\begin{equation}
		|J| = 1+\partial_{||}u_{||}(\mathbf{r}),		\label{eq:jacob}
	\end{equation}
where $\partial_{||}=\partial/\partial r_{||}$ and $u_{||}(\mathbf{r})=\mathbf{u}(\mathbf{r})\cdot\hat{\mathbf{e}}_{||}$ is the line-of-sight component of the scaled velocity field $\mathbf{u}(\mathbf{r})$. Equation \eqref{eq:masscon} can thus be written as
			\begin{equation}
				\delta_m^{s}(\mathbf{s})=\left[\delta_m(\mathbf{r})-\partial_{||}u_{||}(\mathbf{r})\right]\left[1+\partial_{||}u_{||}(\mathbf{r})\right]^{-1}.	\label{eq:masscon2}
			\end{equation}
			
	In the linear regime approximation (i.e. small real-space density field $\left|\delta(\mathbf{r})\right|\ll1$, small velocity gradients $\left|\partial_{||}u_{||}\right|\ll1$ and an irrotational velocity field $\mathbf{\nabla}\x\mathbf{u}(\mathbf{r})=0$), equation \eqref{eq:masscon2} becomes
		\begin{equation}
			\delta_m^{s}(\mathbf{s})=\left[\delta_m(\mathbf{r})-\partial_{||}u_{||}(\mathbf{r})\right]\left[1-\partial_{||}u_{||}(\mathbf{r})\right]. \label{eq:masscon3}
		\end{equation}
Neglecting higher order terms in equation \eqref{eq:masscon3} gives
		\begin{equation}
			\delta_m^{s}(\mathbf{s})=\delta_m(\mathbf{r})-\partial_{||}u_{||}(\mathbf{r}).		\label{eq:masscon4}
		\end{equation}
	In the plane parallel approximation, when the scales of perturbations are assumed to be much smaller then their distances from us (i.e. $u_{||}/r_{||}(\mathbf{r})\ll1$), equation \eqref{eq:masscon4} can be written in Fourier space as
		\begin{equation}
			\delta_m^{s}(k,\mu_k)=\left[1+f(z)\mu_k^2\right]\delta_m(k).		\label{eq:kaiser}
		\end{equation}
	with $\mu_k=k_{||}/k$.
In equation \eqref{eq:kaiser} $f$ is the growth rate of structure defined as:
	\[
		f(a) = \frac{d\ln \delta_m(a)}{d\ln a}.		\label{eq:f}
	\]

	Equation \eqref{eq:kaiser} is valid for the overall matter density field. However, when objects of a given type $i$ are used as tracers of the overall matter density field, their distribution is in general biased with respect to the underlying matter. At first order, the overdensity $\delta_i$ of a given population $i$ can be related to that of the overall matter $\delta_m$ through a linear bias factor $b_{i}$
		\begin{equation}
			\delta_i(\mathbf{k})=b_{i}\,\delta_m(\mathbf{k}).		\label{eq:bias1}
		\end{equation}
	
	In this work we assume the linear bias $b_{i}$ to be a scale-independent parameter which is valid with good approximation in regimes that are not affected by small-scale non linearities and the ones due to the Baryon Acoustic Oscillations (BAO). 
	
	We also assume the peculiar velocity field of $i$-type objects to be unbiased with respect to that of the overall matter. This assumption holds on scales much larger than the typical size of virialized structures. Indeed, inside such structures dynamical processes such as dynamical friction and tidal disruption may alter the velocity field of objects $i$ with respect to that of the total matter introducing a further velocity bias \citep{munari13, elia12}. Thus, equation \eqref{eq:kaiser} becomes
		\begin{equation}
			\delta_i^{s}(k,\mu_k)=\delta_i(k)+f(z)\,\mu_k^2\,\delta_m(k).		\label{eq:kaiser1}
		\end{equation}
		
Defining the distortion parameter $\beta_i(z)$, related to the population $i$, as
	\begin{equation}	
		\beta_i(z)=\frac{f(z)}{b_{i}(z)},		\label{eq:beta}
	\end{equation}
and combining it with equation \eqref{eq:kaiser1} and the linear bias relation \eqref{eq:bias1} leads to
		\begin{equation}
			\delta_i^{s}(k,\mu_k)=\left[1+\beta_i(z)\mu_k^2\right]\delta_i(k)	.	\label{eq:kaiser2}
		\end{equation}

Given two population of objects, e.g. individual galaxies \textit{`gal'} and their groups \textit{`gr'}, their cross power spectrum is defined as
		\begin{equation}
			\<\delta_{\rm gal}(\mathbf{k}_1)\delta_{\rm gr}(\mathbf{k}_2)\>=(2\pi)^3\delta_D(\mathbf{k}_2-\mathbf{k}_1)P_{\rm cr}(\mathbf{k}).		\label{eq:powerspec}
		\end{equation}
The linear cross power spectrum in redshift space $P_{\rm cr, lin}^{(s)}(\mathbf{k})$ can be related to that in real space $P_{\rm cr}(\mathbf{k})$ by combining equation \eqref{eq:powerspec} with \eqref{eq:kaiser2}
		\begin{equation}
			P_{\rm cr, lin}^{s}(k,\mu_k)=\left[1+\beta_{\rm gal}\mu_k^2\right]\left[1+\beta_{\rm gr}\mu_k^2\right]P_{\rm cr}(k).		\label{eq:powerspec0}
		\end{equation}
It is useful to write $P_{\rm cr, lin}^{s}(k,\mu_k)$ as a sum of spherical harmonics
		\begin{equation}
			P^{s}_{\rm cr, lin}(k,\mu_k)=\sum_\ell P_{\rm cr, lin}^{s,(\ell)}(k)L_\ell(\mu_k),		\label{eq:powerspec1}
		\end{equation}
here, $L_\ell(\mu_k)$ are the Legendre polynomials and $P_{\rm cr, lin}^{s,(\ell)}(k)$ are the multipole moments of the linear cross power spectrum $P_{\rm cr, lin}^{s}(\mathbf{k})$ given by
		\begin{equation}
			P_{\rm cr, lin}^{s,(\ell)}(k)=\frac{2\ell+1}{2}\int_{-1}^{+1} P_{\rm cr, lin}^{s}(k,\mu_k)L_\ell(\mu_k)d\mu_k.	\label{eq:powerspec1a}
		\end{equation}
		
The equivalent expression for the two-point cross-correlation function $\xi_{\rm cr, lin}^{s}(\mathbf{r})$ is provided by \cite{hamilton92}
	\begin{equation}
		\xi_{\rm cr, lin}^{s}(r_p,\pi)=\sum_\ell \xi_{\rm cr, lin}^{s,(\ell)}(s)L_\ell(\mu).		\label{eq:crosscorr}
	\end{equation}
In equation \eqref{eq:crosscorr}, $r_p$ and $\pi$ are respectively the components of the pair separation $\mathbf{s}$ transverse and parallel to the line of sight, $\mu$ is the cosine of the angle between the pair separation $\mathbf{s}$ and the line of sight and $\xi_{\rm cr, lin}^{s,(\ell)}(s)$ are the multipole moments of $\xi_{\rm cr, lin}^{s}(r_p,\pi)$:
	\begin{equation}
		\xi_{\rm cr, lin}^{s,(\ell)}(s) = i^\ell\int\frac{k^2dk}{2\pi^2}P_{\rm cr, lin}^{s,(\ell)}(k)j_\ell(ks),		\label{eq:mps_pk-to-xi}
	\end{equation}
where $j_\ell(ks)$ are the spherical Bessel functions. The only non-null multipole moments are
	\begin{subequations}
		\begin{align}
			\xi_{\rm cr, lin}^{s,(0)}(s)&=\left[1\!+\frac{1}{3}\!(\!\beta_{\rm gal}+\beta_{\rm gr})+\frac{1}{5}\beta_{\rm gal}\beta_{\rm gr}\right] \xi_{\rm cr}(r)			\label{eq:crosscorr2}\\
			\xi_{\rm cr, lin}^{s,(2)}(s)&=\!\left[\!\frac{2}{3}(\!\beta_{\rm gal}\!+\!\beta_{\rm gr}\!)\!+\!\frac{4}{7}\beta_{\rm gal}\beta_{\rm gr}\!\right]\!\left[\xi_{\rm cr}(r)\!-\!\bar{\xi}_{\rm cr}(r)\right]		\label{eq:crosscorr3}\\
			\xi_{\rm cr, lin}^{s,(4)}(s)&=\!\left[\!\frac{8}{35}\beta_{\rm gal}\beta_{\rm gr}\!\right]\left[\xi_{\rm cr}(r)\!+\!\frac{5}{2}\bar{\xi}_{\rm cr}(r)\!-\!\frac{7}{2}\bar{\bar{\xi}}_{\rm cr}(r)\right],		\label{eq:crosscorr4}
		\end{align}
	\end{subequations}
where $\bar{\xi}_{\rm cr}(r)$ and $\bar{\bar{\xi}}_{\rm cr}(r)$ are the integrals of the real-space angle-averaged two-point cross-correlation function $\xi_{\rm cr}(r)$ \citep{cole94}:
	\begin{subequations}
		\begin{align}
			\bar{\xi}_{\rm cr}(r)&=\frac{3}{r^3}\int_0^r \xi_{\rm cr}(r')r'^2dr'																						\label{eq:integrals0}\\
			\bar{\bar{\xi}}_{\rm cr}(r)&=\frac{5}{r^5}\int_0^r \xi_{\rm cr}(r')r'^4dr'	\, .																				\label{eq:integrals2}
		\end{align}
	\end{subequations}
This model was adopted by \cite{mountrichas09} to measure the bias of the QSOs in 2SLAQ, 2QZ and SDSS, by cross-correlating them with a population of Luminous Red Galaxies (LRG) in 2SLAQ and AAOmega.

	Both $\beta_{\rm gal}$ and $\beta_{\rm gr}$ in fact describe the same growth rate $f\left(z\right)$; therefore using the linear bias relation \eqref{eq:bias1}, the 2PCF of the population $i$, in real space, can be written as
		\begin{equation}
			\xi_i(\mathbf{r})=b_{i}^2\xi_m(\mathbf{r})	.	\label{eq:bias2}
		\end{equation}
	Combining equation \eqref{eq:bias2}, written separately for galaxies $\xi_{\rm gal}$ and groups $\xi_{\rm gr}$, with \eqref{eq:beta} gives
		\begin{equation}	
			\beta_{\rm gr}(z)=b_{12}(z)\beta_{\rm gal}(z)		.  \label{eq:biasratio}
		\end{equation}
	In equation \eqref{eq:biasratio}, $b_{12}(z)$ is the \textit{`relative bias'} between galaxies and groups, defined as
		\begin{equation}
			b_{12}(z)=\left[	\frac{\xi_{\rm gal}}{\xi_{\rm gr}}\right]^{1/2}=\frac{b_{\rm gal}(z)}{b_{\rm gr}(z)}	. \label{eq:bias_ratio}
		\end{equation}
	
This quantity can be estimated directly from the data, once the 
real-space correlation functions of the two populations are measured through projection of
$\xirp$. Defined in this way, $b_{12}$ will be smaller than unity, given the larger
bias of groups with respect to galaxies. We prefer this definition as it
allows us to obtain a more compact notation in the following
equations. Using equation \eqref{eq:biasratio}, therefore, equations \eqref{eq:crosscorr2},\eqref{eq:crosscorr3}, \eqref{eq:crosscorr4} become
		\begin{subequations}
			\begin{align}
				\xi_{\rm cr, lin}^{s,(0)}(s)&=\left[1\!+\frac{1}{3}\beta_{\rm gal}\!(\!1+b_{12})+\frac{1}{5}\beta_{\rm gal}^2b_{12}\right] \xi_{\rm cr}(r)											\label{eq:crosscorr5}\\
				\xi_{\rm cr, lin}^{s,(2)}(s)&=\!\left[\!\frac{2}{3}\beta_{\rm gal}\!(\!1+b_{12})+\!\frac{4}{7}\beta_{\rm gal}^2b_{12}\!\right]\!\left[\xi_{\rm cr}(r)\!-\!\bar{\xi}_{\rm cr}(r)\right]					\label{eq:crosscorr6}\\
				\xi_{\rm cr, lin}^{s,(4)}(s)&=\!\left[\!\frac{8}{35}\beta_{\rm gal}^2b_{12}\!\right]\left[\xi_{\rm cr}(r)\!+\!\frac{5}{2}\bar{\xi}_{\rm cr}(r)\!-\!\frac{7}{2}\bar{\bar{\xi}}_{\rm cr}(r)\right]	\, .			\label{eq:crosscorr7}
			\end{align}
		\end{subequations}

	The linear Kaiser/Hamilton model for the two-point auto-correlation function is recovered just by taking $b_{12}(z)=1$ and replacing the real-space two-point cross-correlation function $\xi_{\rm cr}(r)$ and its integrals $\bar{\xi}_{\rm cr}(r)$, $\bar{\bar{\xi}}_{\rm cr}(r)$ with their counterparts in the auto-correlation case.

	\begin{figure*}
			\centering
				\includegraphics[scale=0.32]{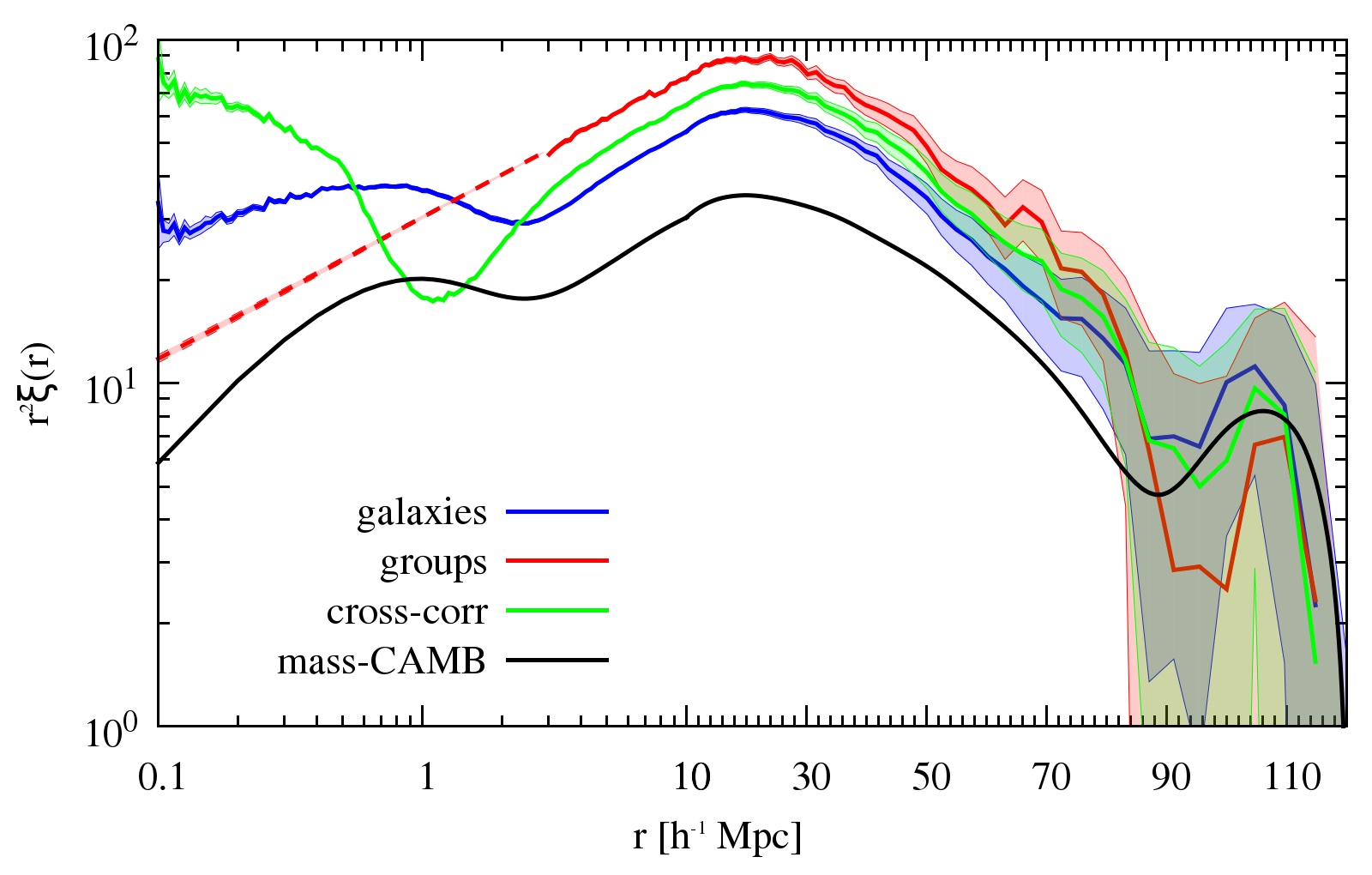}				\caption{\small{Real-space two-point correlation function measurements from the MultiDark Run1 data at $z=0.1$, averaged over 27 equal sub-samples. Blue line shows the (HOD, $L > 1L^{*}$) galaxy auto-correlation, red line represents the auto-correlation of ($M>10^{13}h^{-1}M_\odot$ dark matter haloes) groups while the group-galaxy cross-correlation is plotted with green line. The CAMB model prediction for the correlation function of the underlying matter density field is also plotted (black line). The red dashed line is the power-law extrapolation of the group auto-correlation function on scales below $3\mpcoh$ using parameters in Table \ref{tab:pow-law}. The relative statical errors on the averaged correlation functions are shown as the transparent filled contours.}\label{fig:xir}}
	\end{figure*}

	\subsection{The Dispersion Model}\label{sec:dispersion}
	To derive the linear model in Section \ref{sec:kaiser}, strong assumptions have been made, limiting its validity to the very large scales.
	
	We adopt the mostly used empirical \textit{`Dispersion model'} \citep{peacock94} to model the shape of the 2PCF on small and intermediate trans-linear scales. This model convolves the linear model presented in Section \ref{sec:kaiser} with a peculiar Pairwise Velocity Distribution (PVD) function along the line of sight $\varphi(v_{||})$:
		\begin{equation}
			\xi^{s}(r_p,\pi)=\int_{-\infty}^{+\infty}\xi_{\rm lin}^{s}\left[r_p,\pi-\frac{v_{||}}{aH(a)}\right]\varphi(v_{||})dv_{||}  .		\label{eq:covol}
		\end{equation}
Different functional forms have been proposed in the literature for the PVD function. We adopt an exponential distribution function for $\varphi(v_{||})$ which is found to be in good agreement both with data from N-body simulations and with observations from large galaxy redshift surveys \citep{zurek94}:
		\begin{equation}
			\varphi(v_{||})=\frac{1}{\sqrt{2}\,\sigma_{12}}e^{-\frac{\sqrt{2}\,\left|v_{||}\right|}{\sigma_{12}}}	   .							\label{eq:expPVD}
		\end{equation}
Here, $\sigma_{12}$ is usually referred to as the peculiar pairwise velocity dispersion along the line of sight. In this paper we assume $\sigma_{12}$ to be a scale-independent free parameter. More complicated and accurate models for the peculiar pairwise velocity distribution have been calibrated on simulations to take into account the scale dependence of the velocity distribution \citep[e.g.][]{zu13, bianchi14}.

	\subsection{Truncated Multipole Moments}
	The standard multipole moments of the 2PCF $\xi^{s}(\mathbf{s})$ are obtained by projecting it onto the Legendre polynomials:
		\begin{equation}
			\xil=\frac{2\ell+1}{2}\int_{-1}^{+1}\xi^{s}(s,\mu)L_\ell(\mu)d\mu	,								\label{eq:projmps}
		\end{equation}
	with $s^2=r_p^2+\pi^2$ and $\mu=\pi/s$.
	
	Since strong non-linear distortions affect small transverse $(r_p)$ scales, we propose an alternative way of using the multipole moments of the 2PCF by removing such scales. This means that, at a given scale $s$, we consider only the clustering signal at $r_p>\bar{r}_p$. The new `multipole moments' are then given by
		\begin{equation}
			\xilt=\frac{2\ell+1}{2}\int_{-\bar{\mu}}^{+\bar{\mu}}\xi^{s}(s,\mu)L_\ell(\mu)d\mu ,						\label{eq:projmps_rpcut}
		\end{equation}
where
		\begin{equation}
			\bar{\mu}=\sqrt{1-\(\frac{\bar{r}_p}{s}\)^2}	\, .		\label{eq:mu_bar}
		\end{equation}

It is important to stress here that although $\xilt$ are not, mathematically speaking, the multipole moments of the 2PCF, in the following part, for the sake of simplicity, we will refer to them as the \textit{`truncated multipole moments'}.

The gain in using the truncated multipole moments is given by the fact that they reconcile the classical approach of using the anisotropic 2PCF $\xi(r_p,\pi)$ with the one of using its multipole moments. Indeed, the anisotropic 2PCF $\xi(r_p,\pi)$  allows us to systematically exclude from the fit small transverse $r_p$ scales, affected by strong non-linear distortions, which are difficult to model analytically. However, this approach would involve a huge covariance matrix: it is thus in practice computationally infeasible. The size of the covariance matrix in the case of multipole moments is much smaller, but nonlinearities on small transverse scales make an undesired contribution when projecting the redshift-space 2PCF $\xi^{s}(s,\mu)$ onto the Legendre polynomials. The truncated multipoles allow us to remove these strong non-linear distortions, while retaining a relatively small covariance matrix and yielding a numerically tractable problem.

\section{Methodology}	\label{sec:methodology}																			

To test the model presented in Section \ref{sec:model}, we make use of
simulated catalogues of galaxies and groups based on the MultiDark
Run1 (MDR1) dark matter N-body simulation
\citep{prada12}. MDR1 is a high mass resolution simulation
within a cube of side $1000\mpcoh$, with $N_p=2048^3$ dark matter
particles and the mass of each particle being
$M_p=8.721\times10^9h^{-1}M_\odot$. It assumes a $\Lambda$CDM
cosmology using cosmological parameters from WMAP5 and WMAP7 data
releases
$(\Omega_m,\Omega_\Lambda,\Omega_b,n_s,h,\sigma_8)=(0.27,0.73,0.0469,0.95,0.7,0.82)$. Dark
matter haloes in MDR1 are identified using the Friends-of-Friends
(FoF) algorithm.  

We use the simulated galaxy catalogue used in
\citejap{delatorre12}, which was constructed by populating
MDR1 dark matter haloes and specifying the Halo Occupation
Distribution (HOD). It contains 2,945,687 galaxies with a B-band
luminosity of $L>L^*$. We used the HOD parametrization of
\cite{zheng2005} with parameters $(\log M_{min},\sigma_{\log m},\log
M_0,\log M_1,\alpha)=(12.18,0.21,12.18,13.31,1.08)$ with masses given
in units of $M_\odot/h$. The details of the HOD model that has been
used and the construction of the sample are provided in Appendix B
of \citejap{delatorre12}.  We only remark here that halos are assumed
  isotropic and spherical, with satellite galaxy velocities drawn from Gaussian
  distribution functions along each Cartesian direction, with 
  dispersion computed following \cite{vandenbosch04}.

For the group catalogue, we use all
dark matter haloes with mass $M>10^{13}h^{-1}M_\odot$ giving a total
of 350,518 groups. This specific mass
  threshold is chosen as to match the observed number density
  of groups in the 2dFGRS Percolation-Inferred Galaxy
  Group (2PIGG) catalogue drawn from the 2dFGRS \citep{eke04}.

\subsection{Estimator}
We measure the 2PCF using the minimum variance estimator proposed by \cite{szalay93} adapted to the case of cross-correlation function:
	\begin{equation}
		\xi_{LS}(\mathbf{s}) = \frac{D_1D_2(\mathbf{s})-D_1R(\mathbf{s})-D_2R(\mathbf{s})+RR(\mathbf{s})}{RR(\mathbf{s})},		\label{eq:LS}
	\end{equation}
where $D_1D_2$ are the $data-data$ pairs between the two different catalogues, $D_iR$ are the $data_i-random$ pairs and $RR$ are the $random-random$ pairs. All pair counts are normalized to the related total numbers of distinct pairs.
	
Multipole moments of the measured two-point correlation function $\xi(\mathbf{s})$ are obtained by simply projecting it onto the Legendre polynomials (equation \ref{eq:mps_estim}):

	\begin{widetext}
		\begin{equation}
                  \xil=\frac{2\ell+1}{2}\int_{-1}^{+1}\left[\frac{D_1D_2(s,\mu)-D_1R(s,\mu)-D_2R(s,\mu)+RR(s,\mu)}{RR(s,\mu)}\right]L_\ell(\mu)d\mu		\label{eq:mps_estim}
		\end{equation}
	\end{widetext}

To measure the truncated multipole moments, containing the clustering signal only on transverse scales $r_p>\bar{r}_p$, the integration in equation \eqref{eq:mps_estim} is truncated at $\left[-\bar{\mu},+\bar{\mu}\right]$ rather than $\left[-1,+1\right]$ with $\bar{\mu}$ given by equation \eqref{eq:mu_bar}. We make use of a random catalogue with a number of random points 10 times larger than the galaxy sample to reduce the shot noise in our measurements.
	
	\begin{figure*}
		\centering
			\subfigure[Standard multipoles]{\label{fig:mps_meas}\includegraphics[scale=0.42]{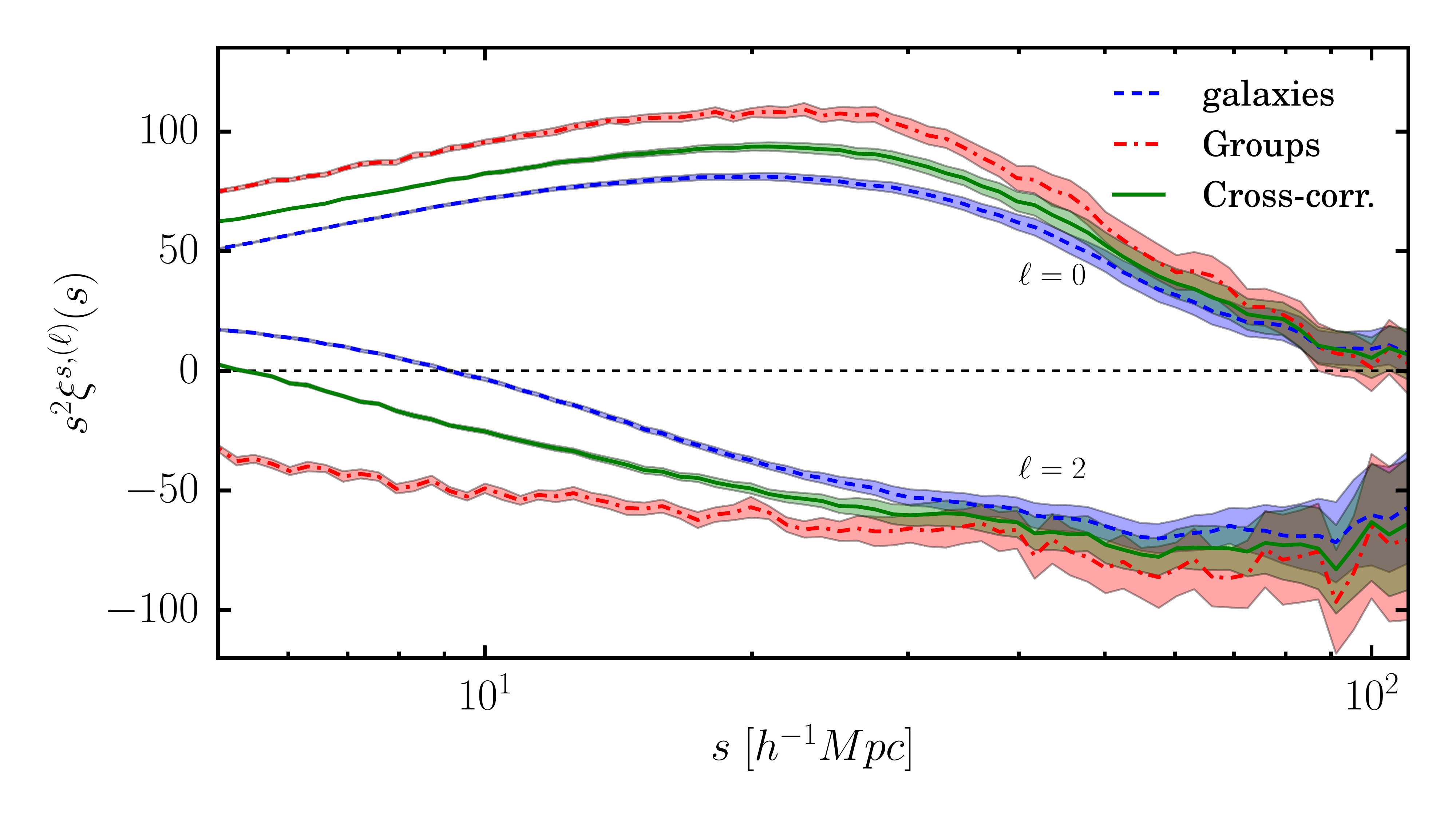}}
  			\subfigure[Truncated multipoles]{\label{fig:mps-rp_meas}\includegraphics[scale=0.42]{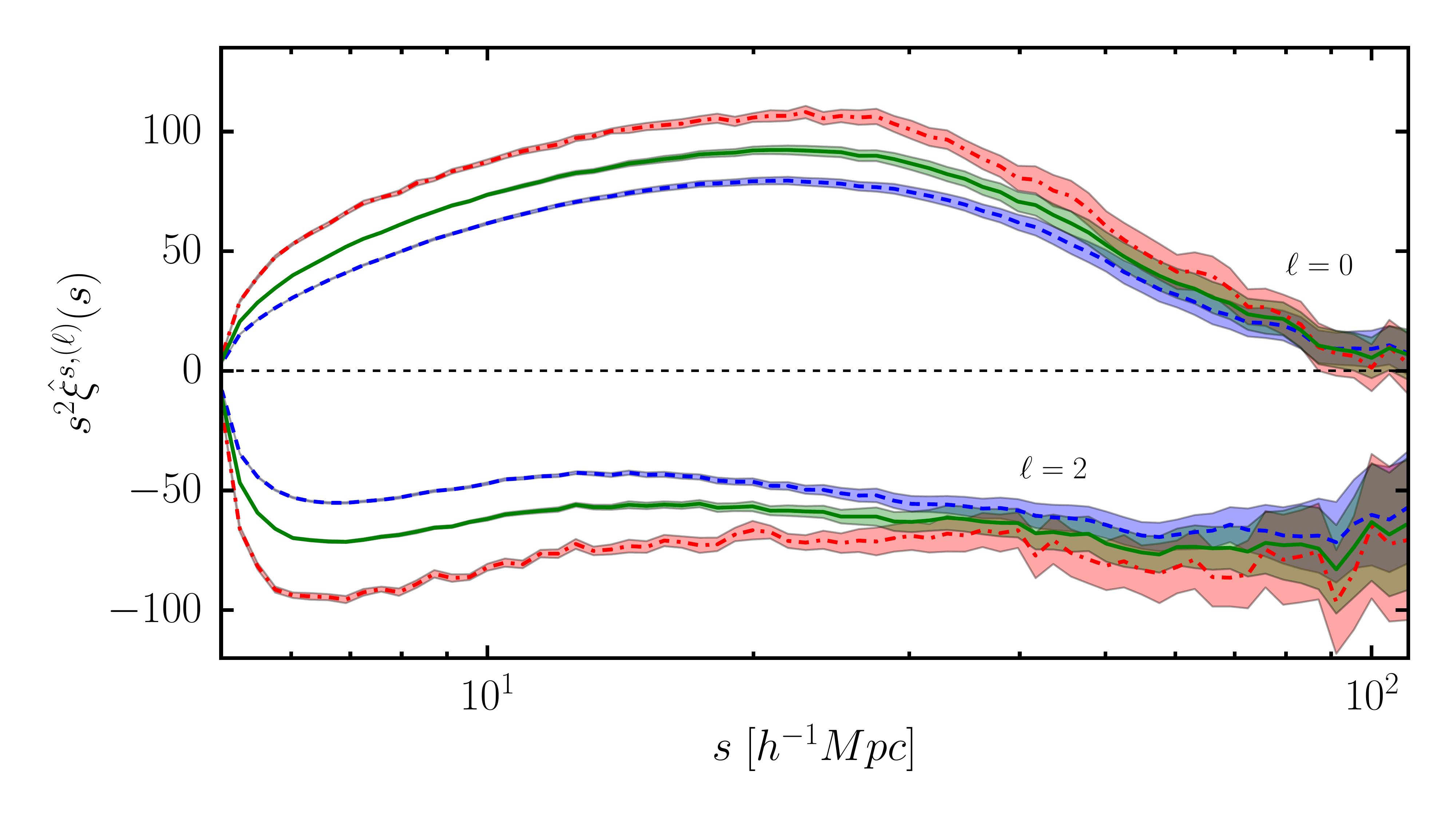}}
			\caption{\small{Standard (top panel) and truncated (bottom panel) monopole $\ell=0$ and quadrupole $\ell=2$ of the 2PCF in redshift space averaged among $N_s=27$ equal sub-volumes drawn from the MDR1 simulation at $z=0.1$. The blue dashed lines show the auto correlation of (HOD, $L>1L^{*}$ ) galaxies, red dash-dotted lines represent the group ($M>10^{13}h^{-1}M_\odot$ dark matter haloes) auto-correlation and their cross correlation is plotted with green continuous lines. The shaded regions show the related rms among $N_s$ measurements scaled by $1/N_s$.}\label{fig:mps_all_meas}}
	\end{figure*}
	
	\begin{figure*}
		\centering
			\subfigure[Galaxies (HOD, $L>L*$)]{\label{fig:xirppi_gal_meas}\includegraphics[scale=0.31]{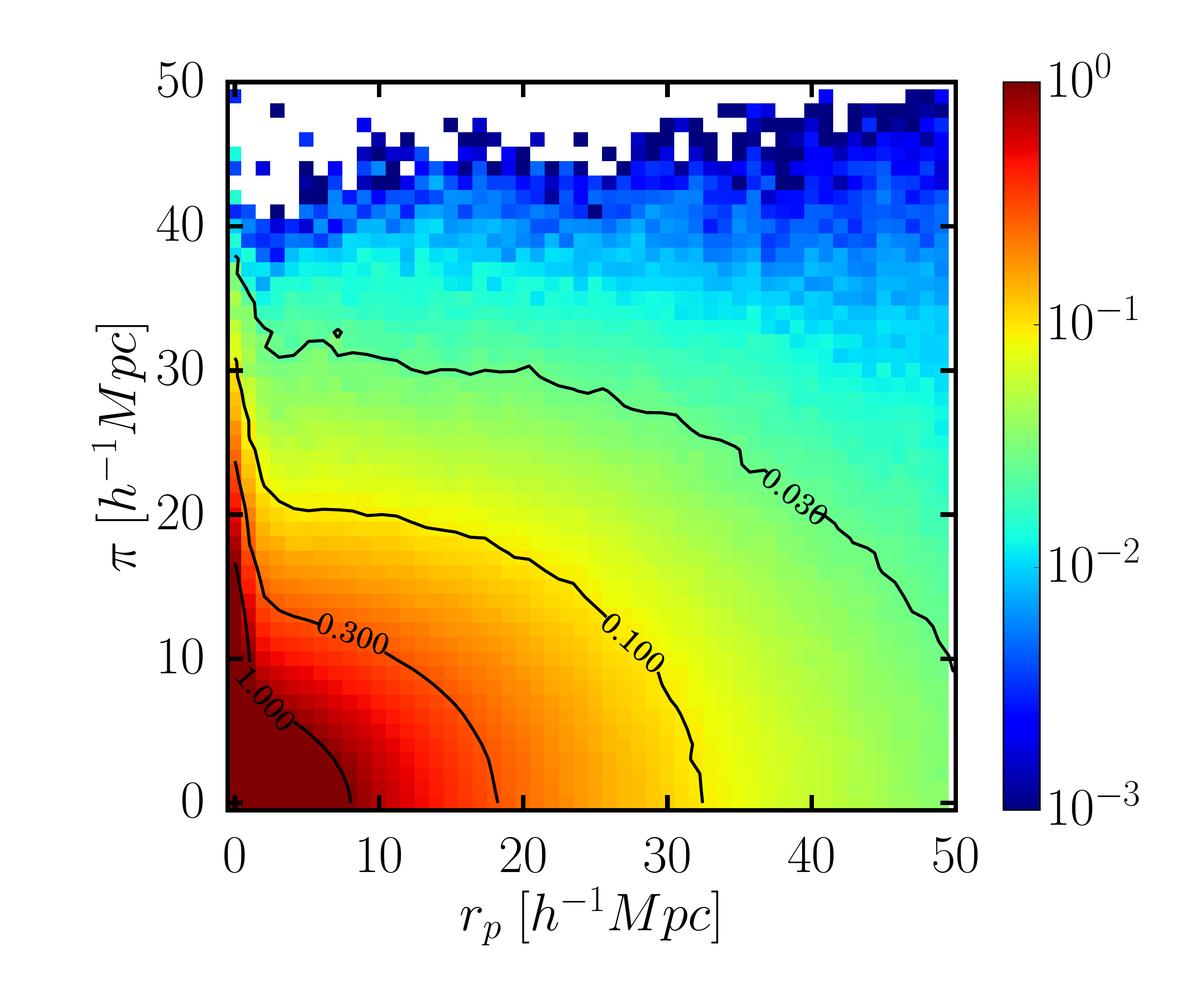}}
  			\subfigure[Groups (Haloes $M>10^{13}h^{-1}M_\odot$)]{\label{fig:xirppi_gr_meas}\includegraphics[scale=0.31]{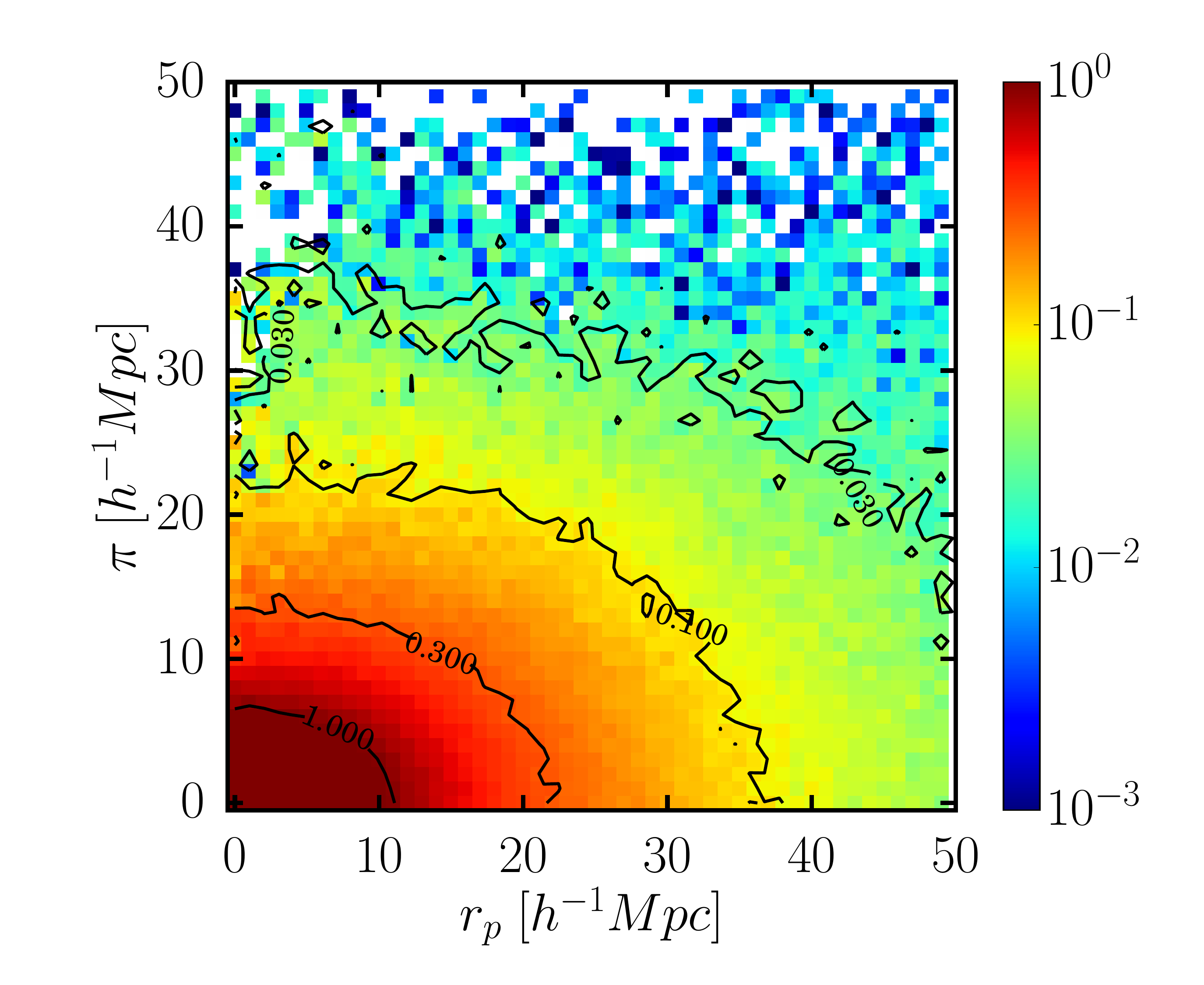}}
			\subfigure[Cross-correlation]{\label{fig:xirppi_cc_meas}\includegraphics[scale=0.31]{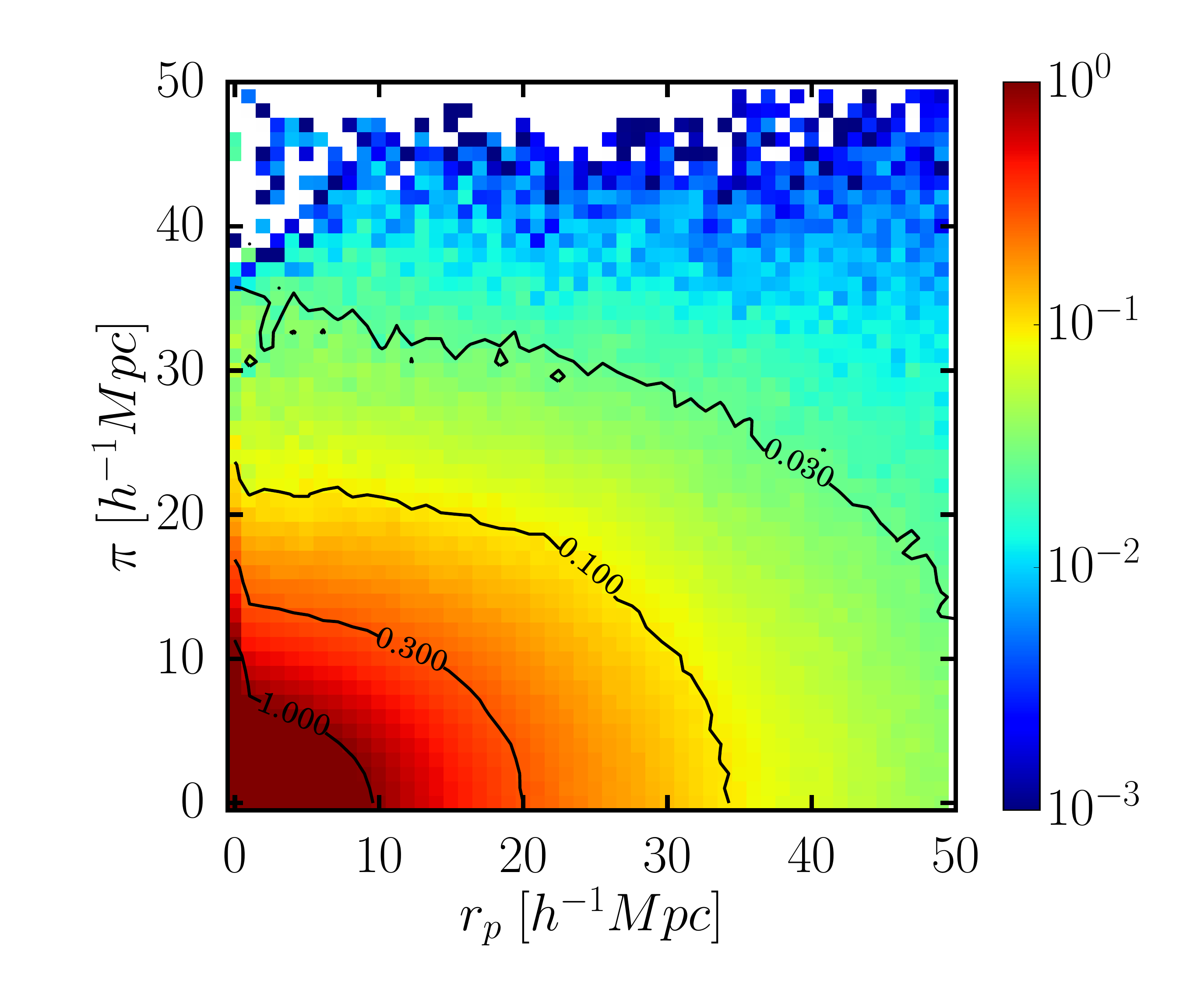}}
			\caption{\small{Anisotropic two-point auto-correlation functions $\xi^s(r_p,\pi)$ for galaxies (top left panel), groups (top right panel) and their cross-correlation (bottom panel) averaged among $N_s=27$ equal sub-volumes.}\label{fig:xirppi_meas}}
	\end{figure*}

We use logarithmic $s$ bins of size $\Delta s_{log}=0.02$ that
  covers the range $\left[0.1, 200\right]\mpcoh$. $\mu$ is divided
  into $200$ linear bins between $\left[0,1\right]$. The measurements
  of the real-space correlation functions are shown in Figure
  \ref{fig:xir}, while the measured standard and truncated multipoles
  for the case of $\bar{r}_{p}=5\mpcoh$ are shown in Figure
  \ref{fig:mps_all_meas}. For the anisotropic 2PCF $\xi(r_p,\pi)$,
  both transverse $r_p$ and parallel to the line-of-sight $\pi$
  components of pair separation are linearly binned with bins of size
  $1\mpcoh$ in the interval $\left[0,100\right]\mpcoh$. The measured
  anisotropic 2PCF are presented in Figure \ref{fig:xirppi_meas}.
	\begin{equation}
		\log_{10}(r_{i+1})=\log_{10}(r_i)+s_{log}		\label{eq:log_bin}
	\end{equation}
Measurements in linear bins are sampled at the mid point of each bin while in the case of logarithmic binning a logarithmic average (eq. \ref{eq:log_av}) of the two edges of each bin is taken as the reference point.
	\begin{equation}
		\log_{10} \langle r_i\rangle = \frac{\log_{10} r_i+\log_{10} r_{i+1}}{2}	\label{eq:log_av}
	\end{equation}
In the last part of the analysis, we fix the number of bins in $s$ to $N_b=12$ inside the range $[s_{min},80]h^{-1}$Mpc in order to fit multipole moments with the full covariance matrix (see Section \ref{sec:covariance}).

\subsection{Error Covariance Matrix}\label{sec:covariance}
	\begin{figure*}
		\centering
			\subfigure[Covariance matrix]{\label{fig:orig}\includegraphics[scale=0.31]{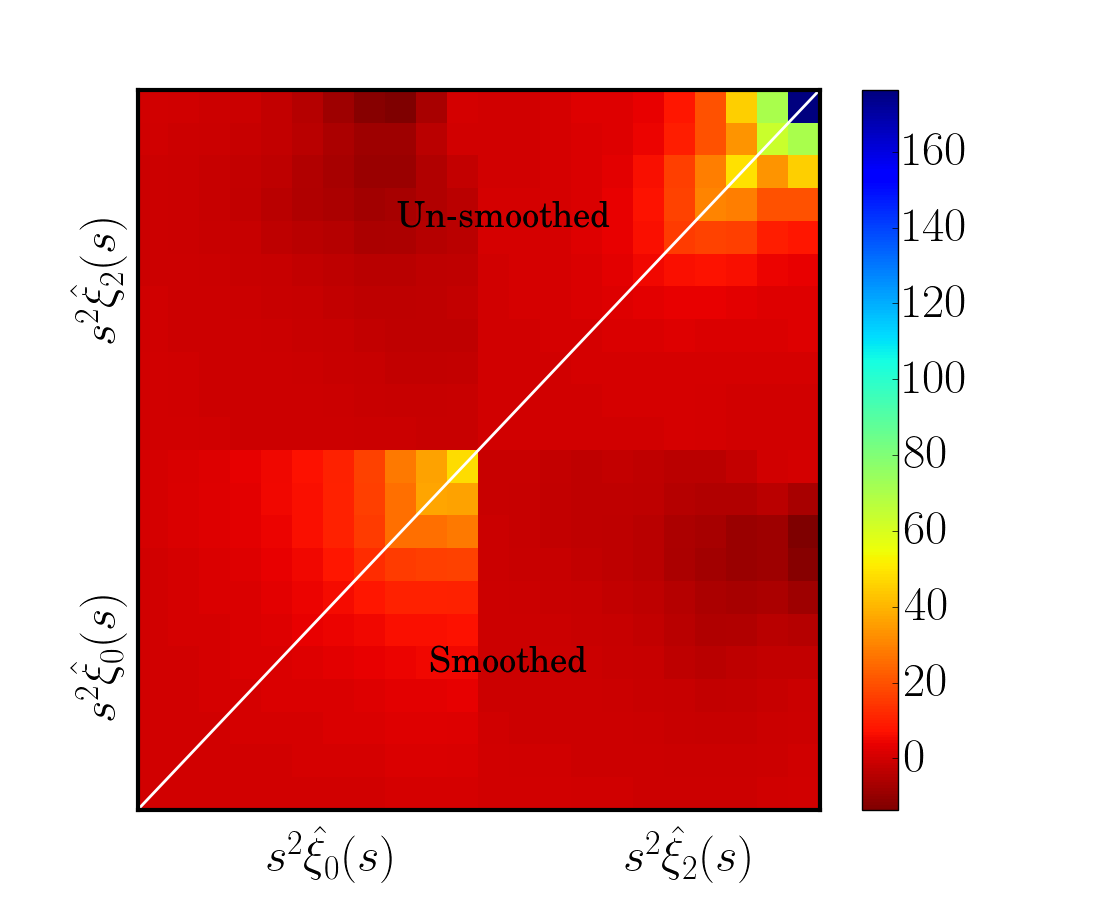}}
			\subfigure[Correlation matrix]{\label{fig:smooth}\includegraphics[scale=0.31]{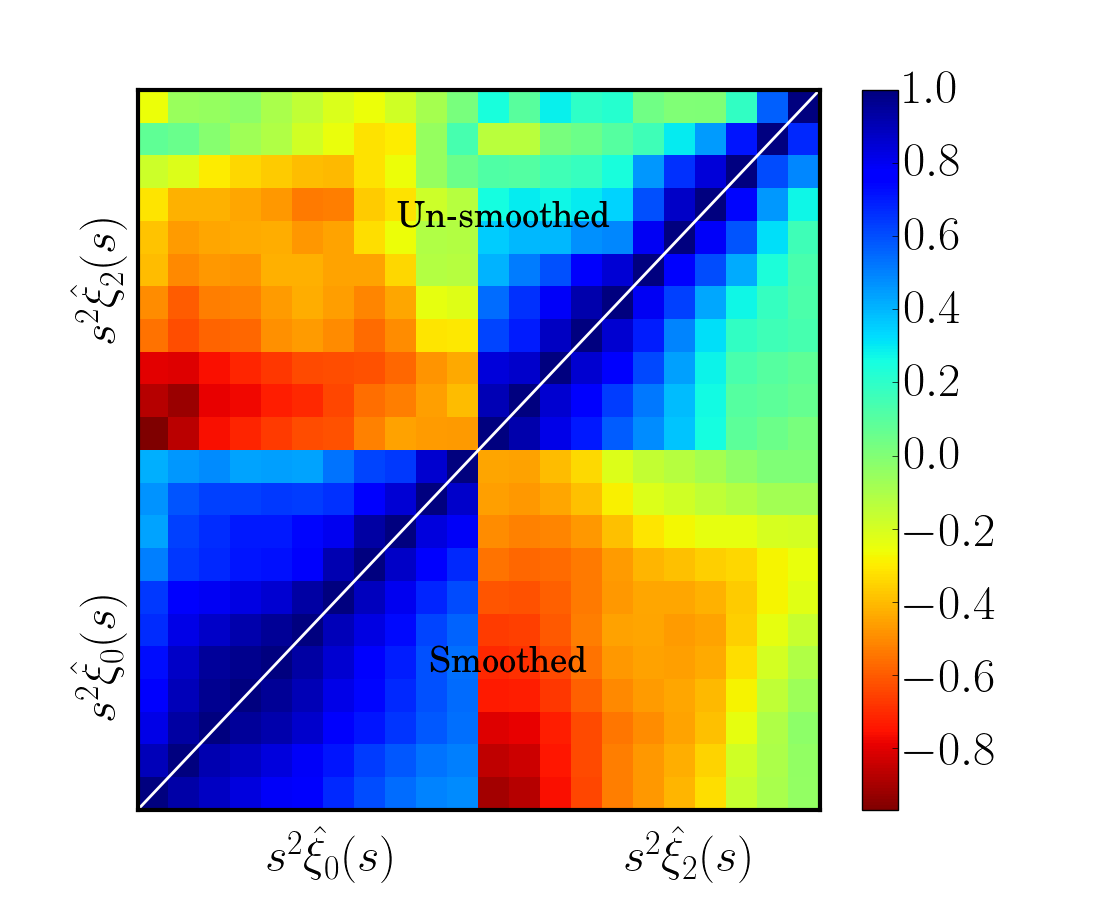}}
			\caption{Covariance (left panel) and correlation (right panel) matrices of the truncated multipoles of the two-point group-galaxy cross-correlation between $[5,80]h^{-1}$Mpc. In each panel the matrix above the white continuous line shows the direct measurement from $27$ equal sub-volumes while the lower part show the matrix after it has been smoothed using a box-car algorithm on $3\times3$ sub-matrices (right panel).}\label{fig:smooth_cij}
	\end{figure*}

In principle, the determination of the covariance matrix requires many
independent realizations of our dataset, of which we inevitably
possess only a single example.  As an approximation to the ideal case,
essentially two classes of methods have been proposed: (a) use
simulated data, to produce $N_r$ independent mock samples with
properties as close as possible to the observed data; (b) use
\textit{internal methods}, in which $N_r$ multiple realizations are
constructed from the overall data, by resampling the observations in
some appropriate way. In particular, four such methods have been used
in the past literature: (1) the classical \textit{bootstrap method},
in which each realization is constructed by resampling with
replacement of single objects in the data set. In the other three
methods, the sample volume is split into a number of sub-volumes,
which are then combined following different schemes: (2) the
\textit{block-wise bootstrap method} builds realizations by combining
all sub-volumes, but assigning a random weight to each of them, to
test the sensitivity of the measured statistics with respect to
specific parts of the sample; (3) the \textit{jack-knife method}
builds realizations simply obtained by omitting one of the
sub-volumes; finally (4) the \textit{sub-sample method} treats each
sub-volume as a realization of the available dataset, yielding a
single estimate of the set of the statistics under evaluation.

Of these internal methods, the sub-sample method is distinct, as it explicitly considers datasets whose volumes are lower than the original parent sample; conversely, the bootstrap and jack-knife methods attempt to estimate a covariance matrix that is appropriate for the whole sample. However, if we are willing to work with smaller volumes, the sub-sample method is closer to the ideal case of many realizations. The sub-samples are not truly independent, and so the results will not be correct on scales approaching that of the sub-volume, but otherwise the sub-sample method should yield directly a reliable covariance matrix for samples having the size of a single sub-volume. If we divide the initial dataset into $N_s$ sub-volumes, we can derive the desired statistic (anisotropic 2PCF or its multipoles, in this case) for each sub-volume, so the natural thing to do is to average these sub-estimates (as opposed to estimating the statistic over the whole sample). In the small-scale limit where the sub-samples can be treated as independent, the covariance matrix for this mean statistic would then be just reduced by a factor $N_s$, since covariances of independent samples add linearly. The overall covariance matrix for the 2PCF, measured as an average over sub-samples, is therefore
	\begin{equation}
		C_{ij}=\frac{1}{N_s}\left\{\frac{1}{N_s-1}\sum_{k}\left[y_i^k-\langle{y}_i\rangle\right]\left[y_j^k-\langle{y}_j\rangle\right]\right\}\, .		\label{eq:cov_sv}
	\end{equation}
Here, $y_i^k$ is the measurement of the statistic $y$ in bin $i$ made from the $k\text{-}th$ realization while $\langle{y}_i\rangle$ is the mean value of $y$ in the same bin over the $N_s$ realizations. We emphasise again that the quantity $\langle{y}_i\rangle$ is to be used as the result for analysis, and it is not identical to the same statistic evaluated once over the whole volume.
In practice, we divided the dataset into $N_s=3^3$ sub-cubes, so that the side of a single sub-cube is $333\mpcoh$. This is large enough that the scales of interest for RSD (tens of Mpc) can be treated as independent in each sub-volume.

Clustering measurements show strong bin-to-bin correlations
  that needs to be taken into account in the fitting procedure using
  the covariance matrix. Because the anisotropic 2PCF has a large
  number of separation bins, the measurement of a proper covariance
  matrix requires a huge number of independent realizations which is
  not available here. Therefore, in order to fairly compare $\beta$
  parameters obtained from the anisotropic 2PCF and its multipole
  moments, we first restricted the analysis to using diagonal
  errors. We then used the full covariance matrix but only to compare
  results obtained from multipole moments. The multipoles have a
  smaller number of bins which makes easier the estimate of the full
  covariance matrix given a limited number of realisations. In
  general, the number of bins in the multipole moments of the 2PCF
  that can be used should be smaller than the number of independent
  realisations. Indeed, a higher number of bins would inevitably yield
  to a singular covariance matrix \citep[e.g.][]{hartlap07}
  which cannot be inverted. To estimate the covariance matrix we thus
  fixed the number of bins to be $N_b=12$ on each multipole,
  independently of the fitting range we consider. In the case of the
  truncated multipoles, however, the measurement in the first bin is
  always equal to zero. To overcome this issue we exclude the first
  bin from the fitting procedure and estimate the covariance matrix on
  the remaining bins resulting in $N_b=11$.

Given the small number of independent realizations ($N_s=27$)
  that we possess and that is of same order as the number of bins in
  our measurements, the raw estimate of the covariance matrix is not
  well constrained and rather noisy. Therefore, we adopt the method of
  \citejap{mandelbaum13} to reduce its noise, by smoothing the
  off-diagonal elements of the associated correlation matrix $R_{ij}=C_{ij}/(C_{ii}C_{jj})^{1/2}$
  with a box-car algorithm. For this, we use an optimal box size of
  $3\times3$ that avoids altering the correlation matrix structure. It
  is important to stress that in this procedure, the smoothing has
  been applied separately for each sub-quadrant and the diagonal
  elements have been left unchanged. As an illustration, we show in
  Figure \ref{fig:smooth_cij} both the original and smoothed
  covariance and correlation matrices, in the case of the truncated
  multipoles of the group-galaxy two-point cross-correlation function
  in range $s=[5,80]h^{-1}$Mpc. One can see in this figure how the
  global structure of the original covariance matrix is preserved in
  by smoothing process and the noise is reduced for the very
  off-diagonal elements.
\medskip

\subsection{Fitting Method}

The model we presented in Section \ref{sec:model} depends mainly on two free parameters: the distortion parameter $\beta_{i}$ and the dispersion parameter $\sigma_{12}$\footnote{We will express the dispersion parameter in units of length $\left[h^{-1}\text{Mpc}\right]$.}. We fit the measured two-point statistics in redshift space with their models (Section \ref{sec:model}) by minimizing the quantity in equation \eqref{eq:likelihood}:

	\begin{widetext}
	\begin{equation}
		-2\ln\mathcal{L}(\beta,\sigma)=\chi^2(\beta,\sigma)=\sum_{i,j}\left[y_i^{dat}-y_i^{mod}(\beta,\sigma)\right]\cdot C_{ij}^{-1}\cdot\left[y_j^{dat}-y_j^{mod}(\beta,\sigma)\right]		\label{eq:likelihood}
	\end{equation}
	\end{widetext}
here, $\mathcal{L}$ is the likelihood function, $y_i^{dat}$ and $y_i^{mod}$ are respectively the measurement and the model prediction for the fitted quantity in bin $i$ and $C_{ij}^{-1}$ is the inverse covariance matrix. 

In particular, since our model depends only on two parameters $(\beta,\sigma)$, we explore the parameter space using a grid on $\beta\in\left[0,0.60\right]$ with bins of size $\Delta\beta=10^{-3}$ and $\sigma_{12}\in\left[0,10\right]h^{-1}\text{Mpc}$ with bins of $\Delta\sigma_{12}=0.05h^{-1}\text{Mpc}$.

\subsection{Model Construction}
In this section we present the measurements, from the simulated catalogues, of the main ingredients required to construct the model presented in Section \ref{sec:model}.

\subsubsection{Real-Space Correlation}
				
One such ingredient is the angle-averaged 2PCF in real space $\xi(r)$.
Given our aim here, which is to test the relative performances of
different estimators of the growth rate, we use the result directly
available from the simulation itself. For real data this quantity is
also available in principle, either via the projection of $\xirp$, or
by fitting an assumed model.  Specifically, as noted above, we use the
real-space correlation functions averaged among 27 sub-sample
realizations as the input for the model that we denote simply with
$\xi(r)$ in the following. In Figure \ref{fig:xir} we show the
measurements of such two-point correlation functions for galaxy
auto-correlation (blue line), group auto-correlation (red line) and
for the group-galaxy cross-correlation (green line). The correlation
function for the underlying overall matter density field as predicted
by the CAMB model \citep{lewis00} is also presented (black
line). We plot the quantity $r^2\xi(r)$ rather than simply $\xi(r)$,
to enhance the differences between different correlation
functions. The transparent filled areas represent the statistical
errors on the averaged real-space two-point correlation functions.

	\subsubsection{Power-Law Extrapolation}

        \begin{figure}
	  \centering
	  \includegraphics[scale=0.21]{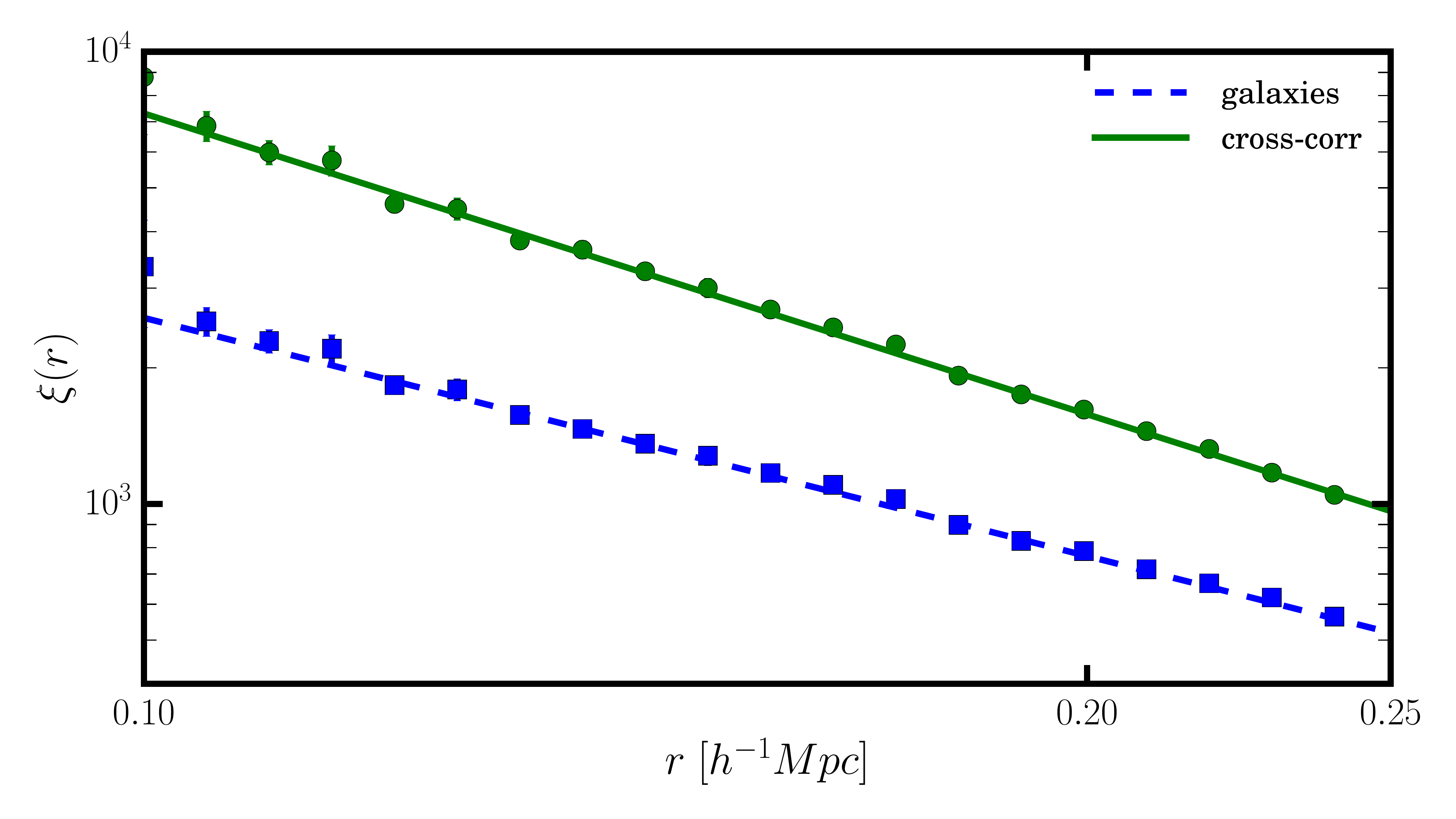}
	  \caption{\small{Power-law fits of the real-space two-point correlation functions, averaged among 27 equal sub-samples, for galaxy auto-correlation (blue dashed line and squares) and the group-galaxy cross-correlation (green continuous lines and circles). Points with error bars represent the measurements while the lines result from the best fit of data using equation \eqref{eq:pow-law} (Table \ref{tab:pow-law}).}\label{fig:pow-law}}
	\end{figure}
        
	\begin{figure}	
	  \centering
	  \includegraphics[scale=0.21]{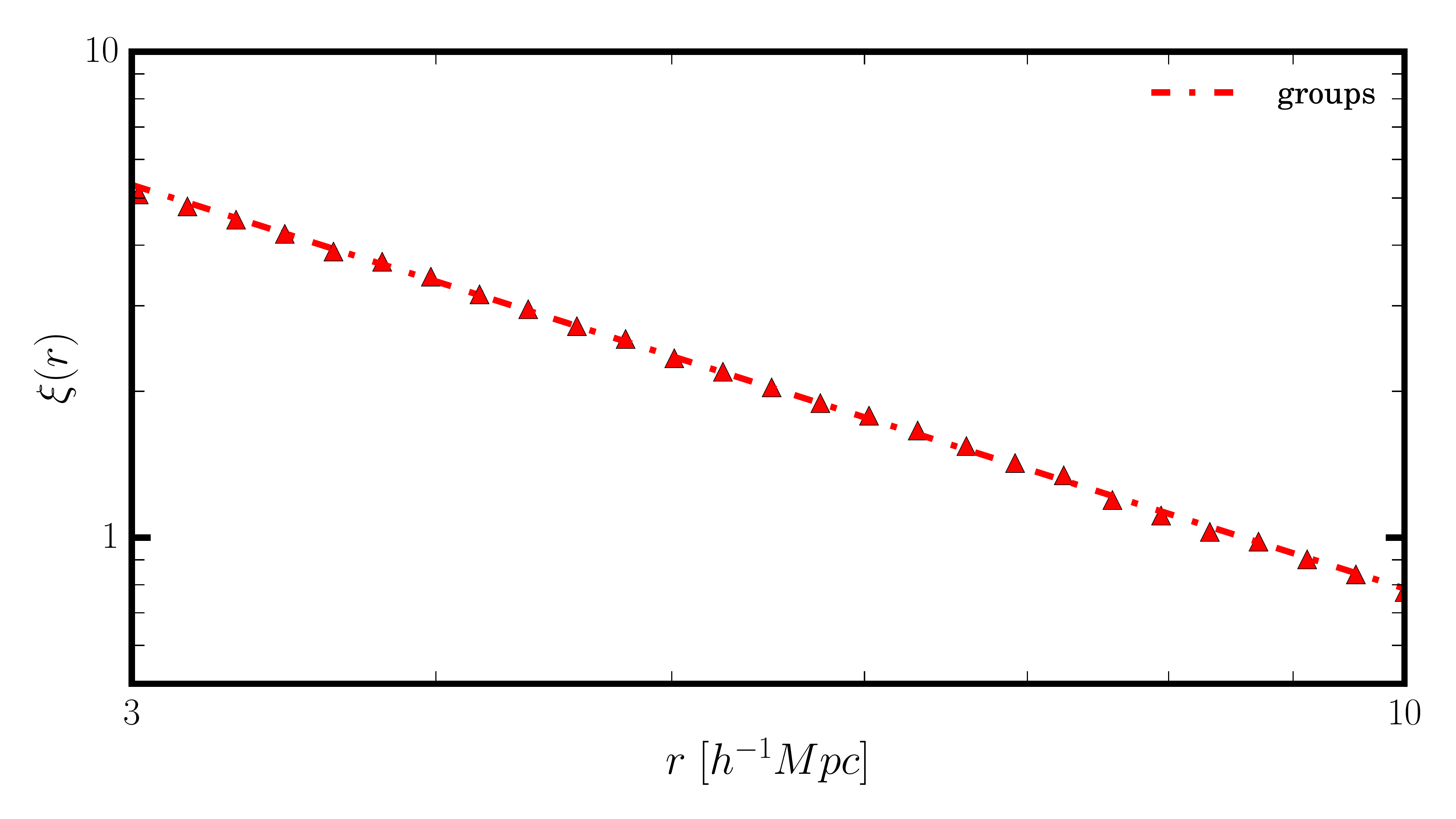}
	  \caption{\small{Same as in Figure \ref{fig:pow-law} but for the case of the group auto-correlation function.}\label{fig:pow-law1}}
	\end{figure}
        
	Although the linear model at a given scale $r$ depends on the integrals of the real-space angle-averaged correlation function $\xi(r)$ between separation zero and $r$, our measurements of $\xi(r)$ are performed within $\left[0.1,200\right]\mpcoh$. We use a power-law form to extrapolate $\xi(r)$ on scales below $0.1\mpcoh$.
		\begin{equation}
			\xi(r<0.1h^{-1}\text{Mpc})=\left(\frac{r}{r_0}\right)^{-\gamma}		\label{eq:pow-law}
		\end{equation}
	The power-law parameters $(r_0,\gamma)$ are measured by fitting $\xi(r)$ on very small scales, i.e. $[0.1,0.25]\mpcoh$ in the case of the galaxy auto-correlation and the group-galaxy cross-correlation functions. For the group auto-correlation, given the low clustering due to the low number pairs of group-sized dark matter haloes on scales below $\sim1\text{-}2h^{-1}$ Mpc, we extrapolate $\xi_{\rm gr}(r)$ on scales below $3h^{-1}\text{Mpc}$ and the power-law parameters are obtained fitting the measured $\xi_{\rm gr}$ in $\left[3,10\right]h^{-1}\text{Mpc}$. The fitting results are listed in Table \ref{tab:pow-law} and shown in Figure \ref{fig:pow-law} for the cross-correlation (green line) and the galaxy auto-correlation(blue line) and in Figure \ref{fig:pow-law1} for the group auto-correlation (red line). In Figures \ref{fig:pow-law} and \ref{fig:pow-law1}, the error bars related to the statistical errors on the measured correlation functions are smaller than the size of the plot symbols.

		 \begin{table}
       			\scriptsize
        			\begin{center}
                			\begin{tabular}{c       c       c}
                        			\hline
					\hline
															&       $r_0$ [$\mpcoh$]					&       $\gamma$					\\
					\hline
                   	     			Galaxy auto-correlation					&	$8.9^{+0.7}_{-0.6}$					&	$1.75^{+0.03}_{-0.03}$			\\
					\hline
						Group auto-correlation					&	$8.59^{+0.02}_{-0.02}$				&	$1.589^{+0.006}_{-0.006}$			\\
					\hline
						Group-galaxy cross-correlation				&	$5.6^{+0.3}_{-0.3}$					&	$2.21^{+0.03}_{-0.03}$			\\
					\hline
					\hline
        				\end{tabular}
      				\caption{\small{Power-law $\xi(r)=(r/r_0)^{-\gamma}$ parameters estimated fitting the real-space two-point auto-correlation function of galaxies and the group-galaxy cross-correlation function, averaged over 27 equal sub-samples, between $[0.1,0.25]\mpcoh$. The same parameters in the case of the group auto-correlation function are estimated fitting it between $[3,10]\mpcoh$}.}\label{tab:pow-law}
	        		\end{center}
		\end{table}

	\subsubsection{$\xi(r)$ Integrals}

        \begin{figure}
	  \centering
	  \includegraphics[scale=0.15]{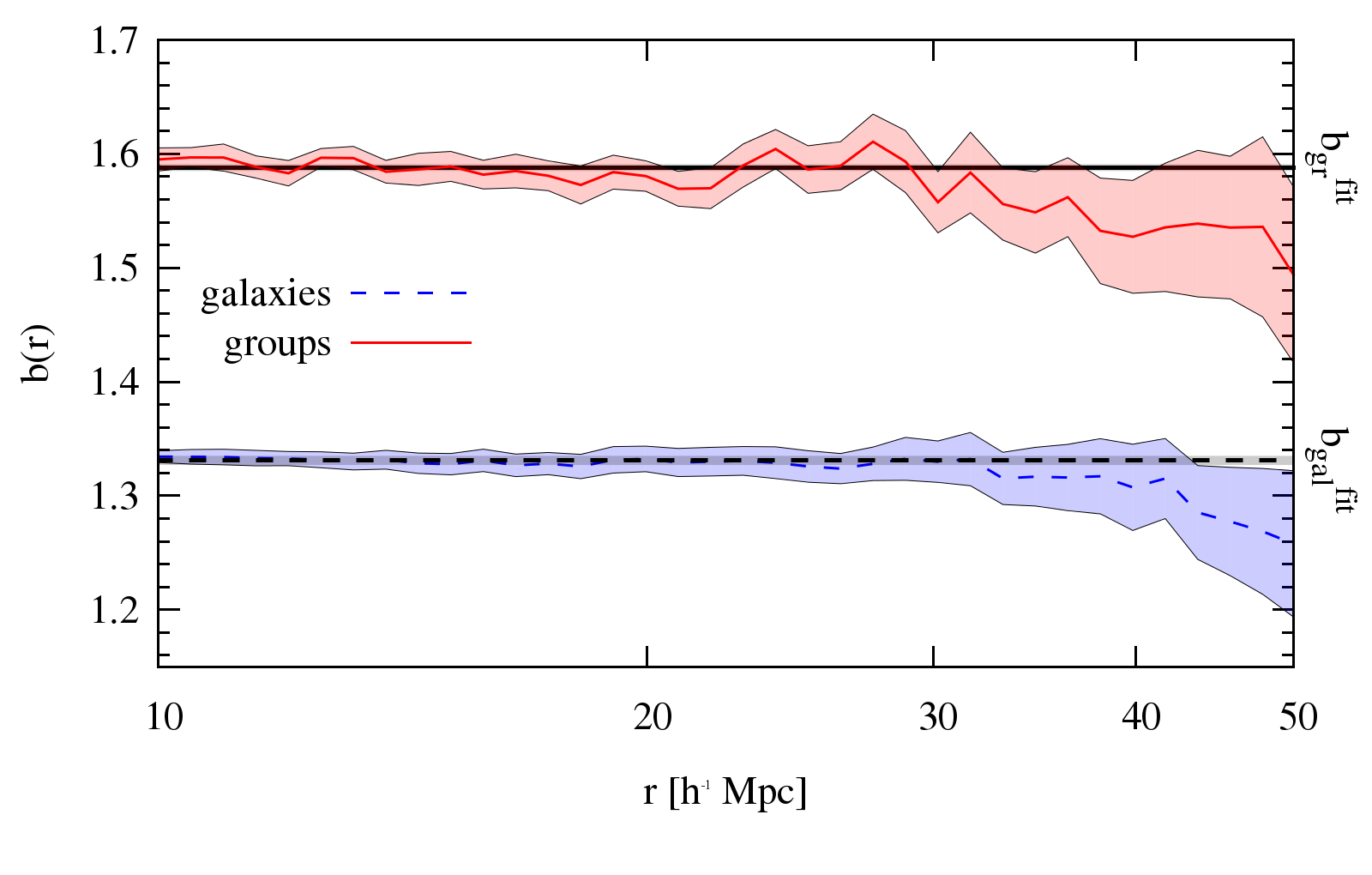}
	  \caption{\small{Measurements of the galaxy linear bias $b_{\rm gal}(r)$ and related $1\text{-}\sigma$ errors are shown with blue dashed line and filled blue contours. The red continuous line with red filled contours represent the measurement of the groups linear bias $b_{\rm gr}(r)$ and its $1\text{-}\sigma$ statistical errors. The black dashed and continuous lines with grey contours represent the best fit of data, respectively for the galaxy and groups linear bias factor, between $\left[10,50\right]\mpcoh$ with a constant model $b^{mod}=\text{const}$.}\label{fig:bias}}
	\end{figure}

\begin{figure*}
		\centering
			\subfigure{\includegraphics[scale=0.21]{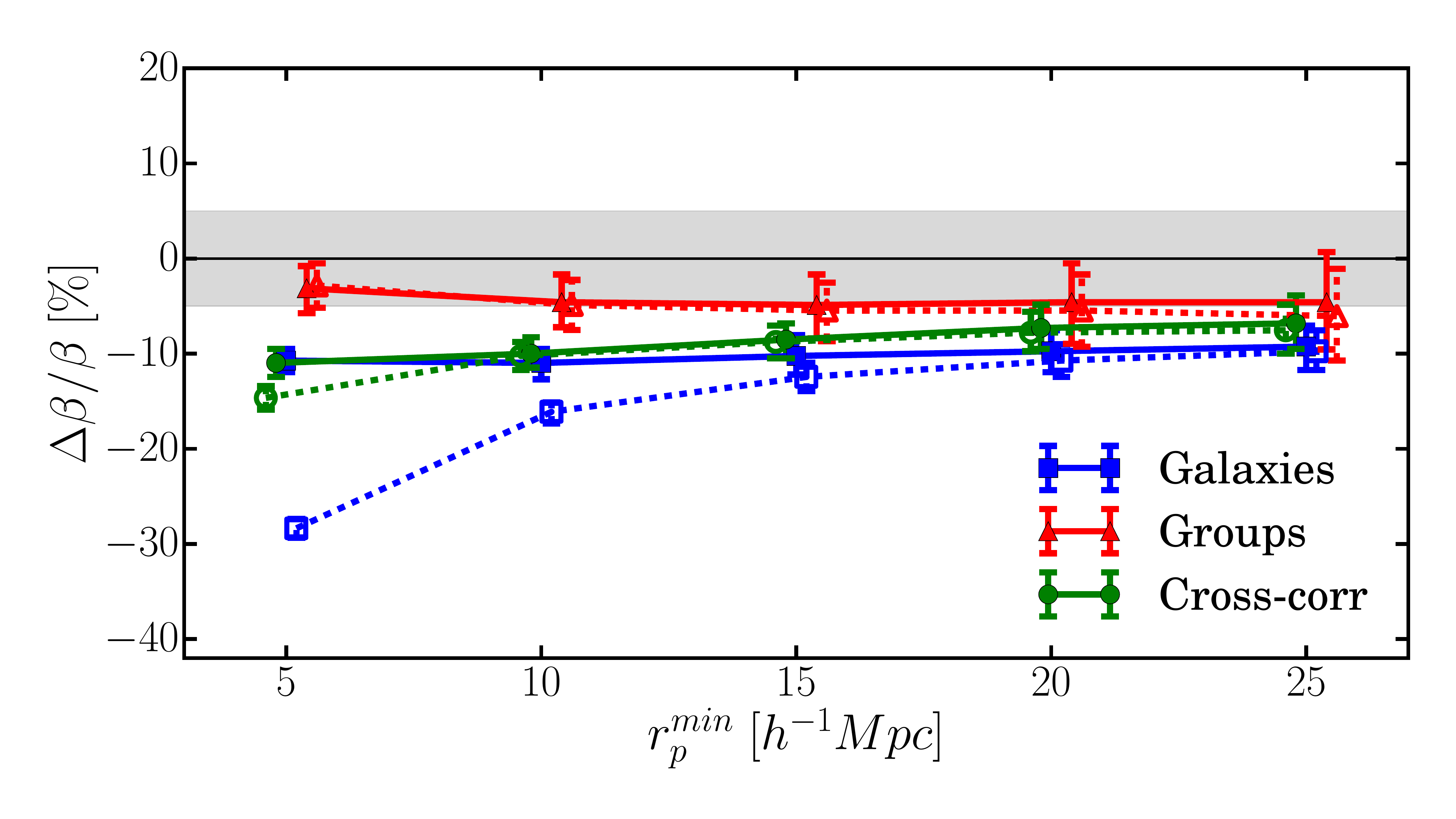}}
			\subfigure{\includegraphics[scale=0.21]{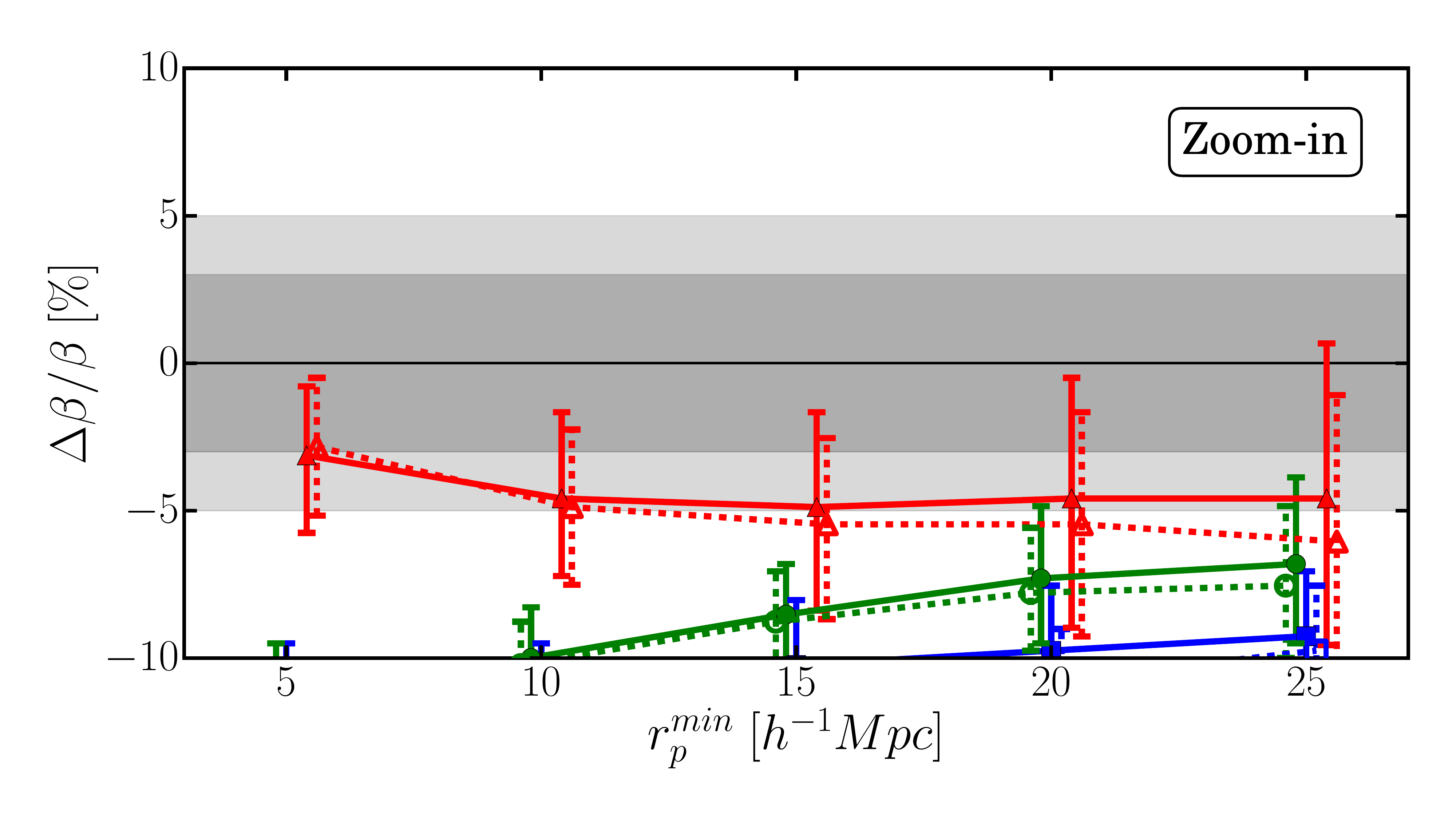}}
			\caption{\small{Systematic errors on the galaxy distortion parameter $\beta_{\rm gal}$ estimated by fitting the anisotropic (HOD, $L>1L*$) galaxy auto-correlation function (blue lines) and the group-galaxy cross-correlation function (green lines) and on the group distortion parameter $\beta_{\rm gr}$ from the anisotropic ($M>10^{13}h^{-1}M_\odot$ dark matter haloes) group auto-correlation function (red lines), using only the diagonal elements of the covariance matrix, are plotted. The continuous lines with filled points represent results using the Dispersion model. The dashed lines with empty points correspond to fits using the linear Kaiser/Hamilton model only. The error bars correspond to the scatter among the 27 sub-samples used in each analysis. The right panel is just a zoom of the left panel. The shaded regions represent the $5\%$ and $10\%$ levels in the left panel while in the right panel are shifted to $3\%$ and $5\%$ level.}\label{fig:ris_rp-pi}} 
\end{figure*}

	Here we compute the integrals of the real-space angle-averaged correlation functions $\xi(r)$ defined in equations \eqref{eq:integrals0} and \eqref{eq:integrals2}. The integration is performed by splitting it into two parts:
		\begin{subequations}
			\begin{align}
				\bar{\xi}(r)=\frac{3}{r^3}\left\{\frac{r_0^\gamma}{3-\gamma}r_{\rm pl}^{3-\gamma}+\int_{r_{\rm pl}}^{r}r'^2\xi(r)dr'	\right\}										\label{eq:xib1}\\
				\bar{\bar{\xi}}(r)=\frac{5}{r^5}\left\{\frac{r_0^\gamma}{5-\gamma}r_{\rm pl}^{5-\gamma}+\int_{r_{\rm pl}}^{r}r'^4\xi(r)dr'	\right\}									\label{eq:xibb1}
			\end{align}
		\end{subequations}
	Here $r_{\rm pl}$ corresponds to the upper limit of the power-law
        extrapolation, which is $r_{\rm pl}=0.1h^{-1}\text{Mpc}$ for the
        galaxy auto-correlation and the group-galaxy cross-correlation
        and $r_{\rm pl}=3h^{-1}\text{Mpc}$ for the group
        auto-correlation. In \eqref{eq:xib1} and \eqref{eq:xibb1} the
        first terms come from the power-laws on scales between
        $\left[0,r_{\rm pl}\right]\mpcoh$, while the second terms are
        the integrals of the effective measurements of the real-space
        2PCFs.

\subsection{Fiducial Model}

Among others, one of the advantages in using data from simulations is
that we know \textit{a priori} the expected (i.e. fiducial) values of
the quantities we want to measure. This makes them an ideal tool to
test and compare the reliability of different methods and theoretical
models. In this part we discuss how the fiducial values of
the galaxy and group distortion parameters, $\beta_{\rm gal}(z=0.1)$ and
$\beta_{\rm gr}(z=0.1)$ have been estimated from the simulated datasets.  

To perform this, we need first to estimate the fiducial value for the linear
bias $b_{i}$ of our galaxies and groups (haloes).  We do this by
fitting the quantity
	\begin{equation}	
		b_{i}(r)=\left[\frac{\xi_{i}(r)}{\xi_{\rm mass}(r)}\right]^{1/2}		\label{eq:bias_1}
	\end{equation}
 in equation
\eqref{eq:bias_1} with a constant between
$\left[10,50\right]\mpcoh$. This choice for the fitting range is
dictated by the need to exclude both small non-linear scales and BAO
scales.  Not having at our disposal the 
catalogue of dark matter particles for the MDR1, we have used
CAMB to predict the matter correlation function
$\xi_{\rm mass}(r)$ at $z=0.1$. The ratio in equation \eqref{eq:bias_1} is shown in Figure \ref{fig:bias} as measured for the galaxies (blue dashed line and contours) and groups (red continuous line and contours). We obtain the following values:
		\begin{subequations}
			\begin{align}
				b_{\rm gal}^{\rm fid}(z=0.1)=(1.331\pm0.004)																\label{eq:bgal}\\
				b_{\rm gr}^{\rm fid}(z=0.1)=(1.588\pm0.003)																\label{eq:bgr}
			\end{align}
		\end{subequations}
The loss of power on scales $>30\mpcoh$ is not problematic since, given the statistical error bars, it is still consistent with the assumption of a constant bias factor.

Once the two values of the linear bias $b_{\rm gal}$ and $b_{\rm gr}$ are measured, the estimation of the relative bias $b_{12}$ is straightforward using eq. \eqref{eq:bias_ratio} and results in:
	\begin{equation}
		b_{12}(z=0.1)= (0.838\pm 0.003)
	\end{equation}
	
Given the cosmology of the MDR1 simulation, we obtain a 
fiducial value $f^{\rm fid}(z=0.1)=0.5434$, corresponding to the
following values for the
distortion parameters $\beta_{\rm gal}$ and $\beta_{\rm gr}$:
		\begin{subequations}
			\begin{align}
				\beta_{\rm gal}^{\rm fid}(z=0.1)=(0.4083\pm 0.0006)\\
				\beta_{\rm gr}^{\rm fid}(z=0.1)=(0.3422\pm 0.0006)
			\end{align}
		\end{subequations}

\section{Results}	\label{sec:results_diag}																				

\begin{figure*}
		\centering
			\subfigure{\includegraphics[scale=0.21]{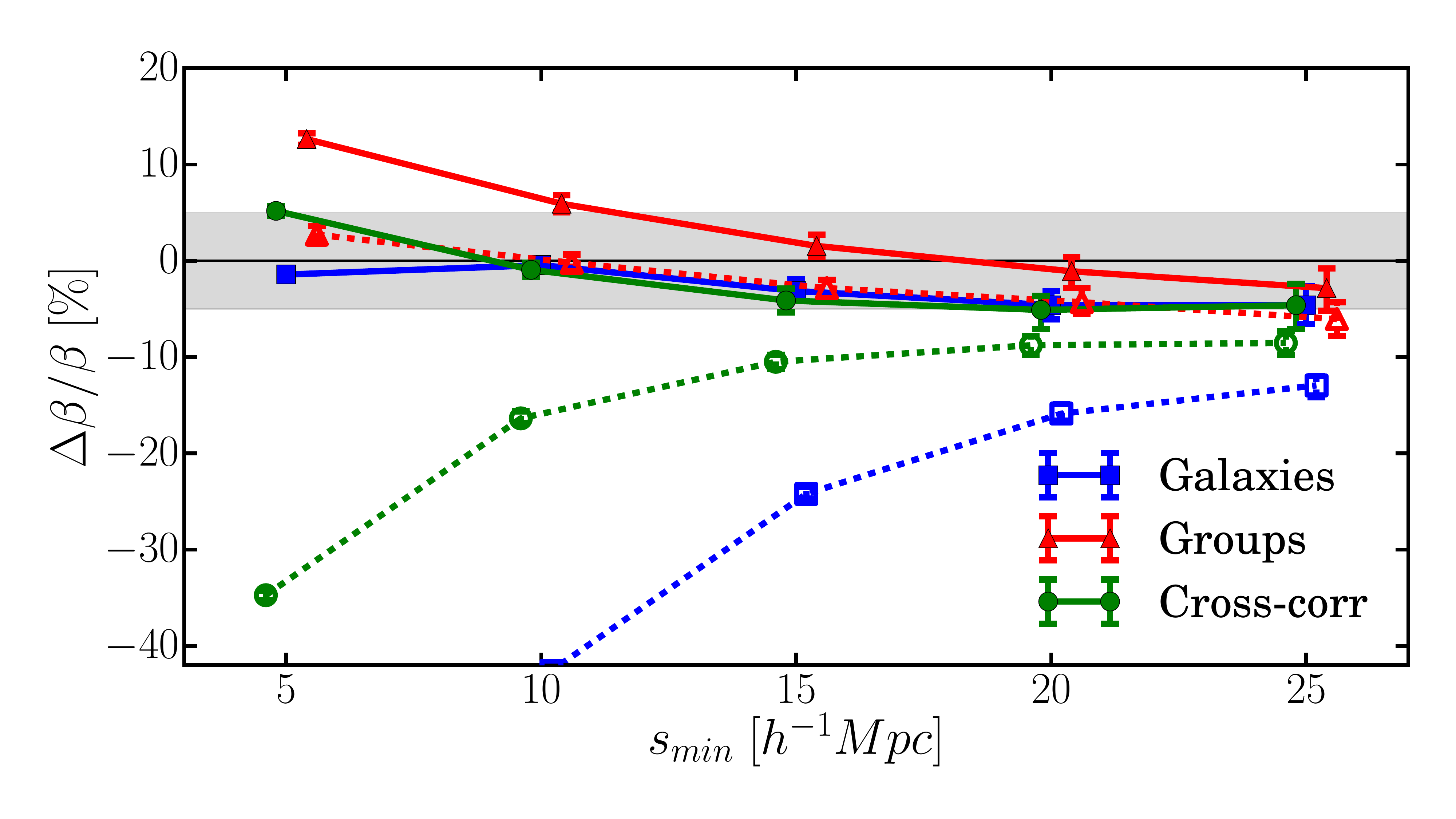}}
			\subfigure{\includegraphics[scale=0.21]{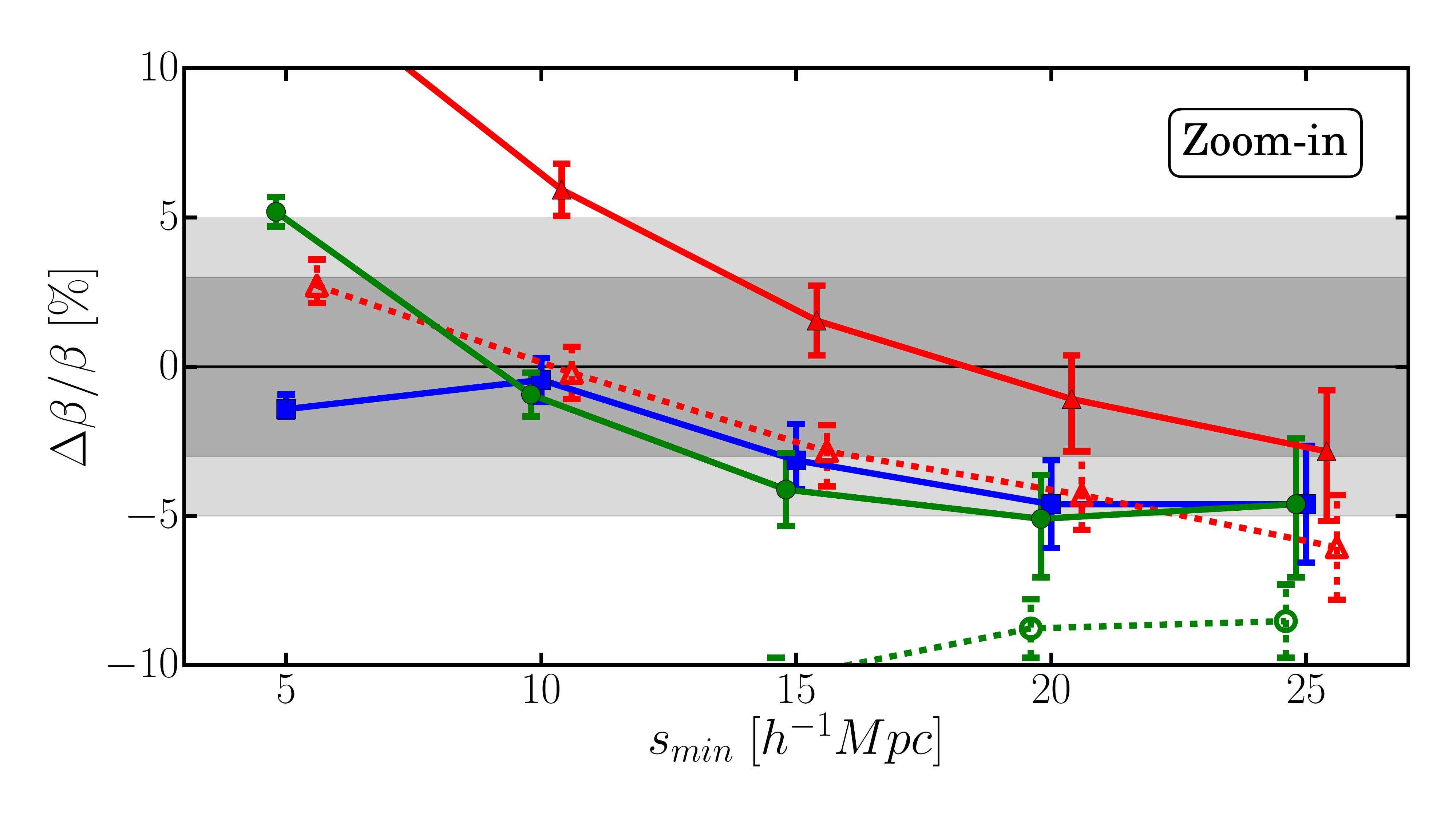}}
			\caption{\small{Same as in Figure
                            \ref{fig:ris_rp-pi} but now fitting
                            monopole and quadrupole moments of the
                            two-point correlation function in redshift
                            space. Continuous and dashed lines
                            correspond respectively to the dispersion
                            and pure linear models. Note how the
                            linear model fails whenever galaxies are
                            involved using either their
                            auto-correlation or the group-galaxy
                            cross-correlation, while it produces a
                            meaningful result when applied to
                            groups.} \label{fig:ris_mps}}
\end{figure*}

We now present the results obtained by fitting the measured two-point correlation
functions with the models presented in Section \ref{sec:model}. We
shall first fit the full $\xi^{s}(r_p,\pi)$, its
multipoles $\xi^{s,(\ell)}(s)$ and the related truncated multipoles
$\hat{\xi}^{s,(\ell)}(s)$ using only the diagonal elements of the
data covariance matrices; we then perform a full covariance analysis
using the truncated multipoles $\hat{\xi}^{s,(\ell)}(s)$.  
 
Several fits are performed, varying the minimum fitting scale in order to
study the impact of non-linearities on the accuracy of the
measurements as a function of scale. In all cases the maximum scale in
the fitting process is limited to a given $s_{\rm max}$, to avoid the
complication of modelling the BAO signature. 

In the case of fitting the linear Kaiser/Hamilton model, there is only a single free parameter, $\beta$.
For the Dispersion model, we must also deal with a nuisance parameter 
in the form of the pairwise dispersion, $\sigma_{12}$. The constraints on $\beta$ that result
from this model are marginalized by integrating the likelihood over a range of $\sigma_{12}$,
assuming a uniform prior within a maximum of $\sigma_{12}=10\mpcoh$. To some extent, there
is a degeneracy between this parameter and $\beta$: raising $\beta$ flattens the contours of $\xi$,
whereas the Fingers of God oppose this tendency. But in practice this degeneracy is not severe, reflected in the fact that the errors on $\beta$ are not much greater for the Dispersion model
than for the linear model. The significance of treating Fingers of God is therefore more in helping
to  reduce bias in the best-fitting value of $\beta$.

\subsection{Fits to the full anisotropic correlation function}

We show here the results of fitting the full $\xi^s(r_p,\pi)$. Fits are
performed varying the minimum transverse scale $r_p^{\rm min}$, and with a
maximum scale fixed at $r_p^{\rm max}=\pi^{\rm max}=50\mpcoh$.  We maximize
the likelihood function defined in equation \eqref{eq:likelihood},
restricted to the diagonal elements of the covariance
matrix.  Following previous work 
\cite{hawkins03}, \cite{guzzo08},
\cite{bianchi12} we fit the quantity
	\begin{equation}
		y(r_p,\pi)=\ln \left[1+\xi^s(r_p,\pi)\right]	\label{eq:logxi}
	\end{equation}
rather than directly the values of $\xi^s(r_p,\pi)$, in order to
enhance the weight of large, more linear scales. 

The results for the galaxy auto-correlation (blue lines), group-galaxy
cross-correlation (green lines) and the group auto-correlation (red lines) are presented in Figure
\ref{fig:ris_rp-pi} and the related statistical errors are shown with
the error bars.  Continuous lines with filled points correspond to
fits performed using the Dispersion model while the dashed lines show
the results when using the pure linear Kaiser/Hamilton model.

When using the Dispersion model, the galaxy
auto-correlation approach underestimates $\beta_{\rm gal}$ by $\sim 10\%$,
while the group-galaxy cross-correlation gives similar results when
including scales below $10h^{-1}\text{Mpc}$. The recovered values of
$\beta_{\rm gal}$ from these two approaches are compatible with each
other. On the other hand, the group auto-correlation results yield a much more
accurate estimate of the group distortion parameter $\beta_{\rm gr}$,
underestimating it by $2\text{-}4\%$. As expected, given the
different size of  the data sets in these different cases, the
statistical errors increase passing from the galaxy auto-correlation
to the group-galaxy cross-correlation to the group
auto-correlation. Results from the auto-correlations of galaxies and
groups are in full agreement with the previous work of
\cite{okumura11}, \cite{delatorre12} and
\cite{bianchi12}. In particular, Figure 5 in the last of these papers shows how the classical approach underestimates $\beta$ by
$\sim10\%$ for the case of haloes with mass $\sim 10^{12} M_\odot$ and
that this systematic error diminishes for increasing halo masses, being
close to unbiased for $\sim 10^{13} M_\odot$  i.e. for group-sized haloes. 

On scales below $15\mpcoh$, the pure linear model heavily
underestimates the distortion parameter with respect to the dispersion
model when applied to the galaxy auto-correlation. Such discrepancy
decreases using first the group-galaxy cross-correlation and
disappears almost completely in the case of the group
auto-correlation. This shows how the corrections to the linear model
on intermediate quasi-linear scales, made through the dispersion
model, become gradually less important using objects less affected by
the non-linear effects on such scales. 

The results from the analysis in this section match our initial
expectations. Specifically, using objects tracing
higher-mass haloes we are in general less sensitive to the details of non-linear
corrections, as previously shown by 
\cite{bianchi12}.  This is particularly true for the auto-correlation of
groups, where we see that the linear and Dispersion models perform
similarly even when including small 
scales. This is at variance with what we observe for galaxies. The use
of the cross-correlation, in this case, represents a reasonable
compromise between having a larger statistics (thus smaller error
bars), while limiting the systematic errors.  In the next session, we
shall see how a different way of fitting the data can further
ameliorate these results, in particular for the cross-correlation
function.

\begin{figure*}
		\centering
			\subfigure{\includegraphics[scale=0.21]{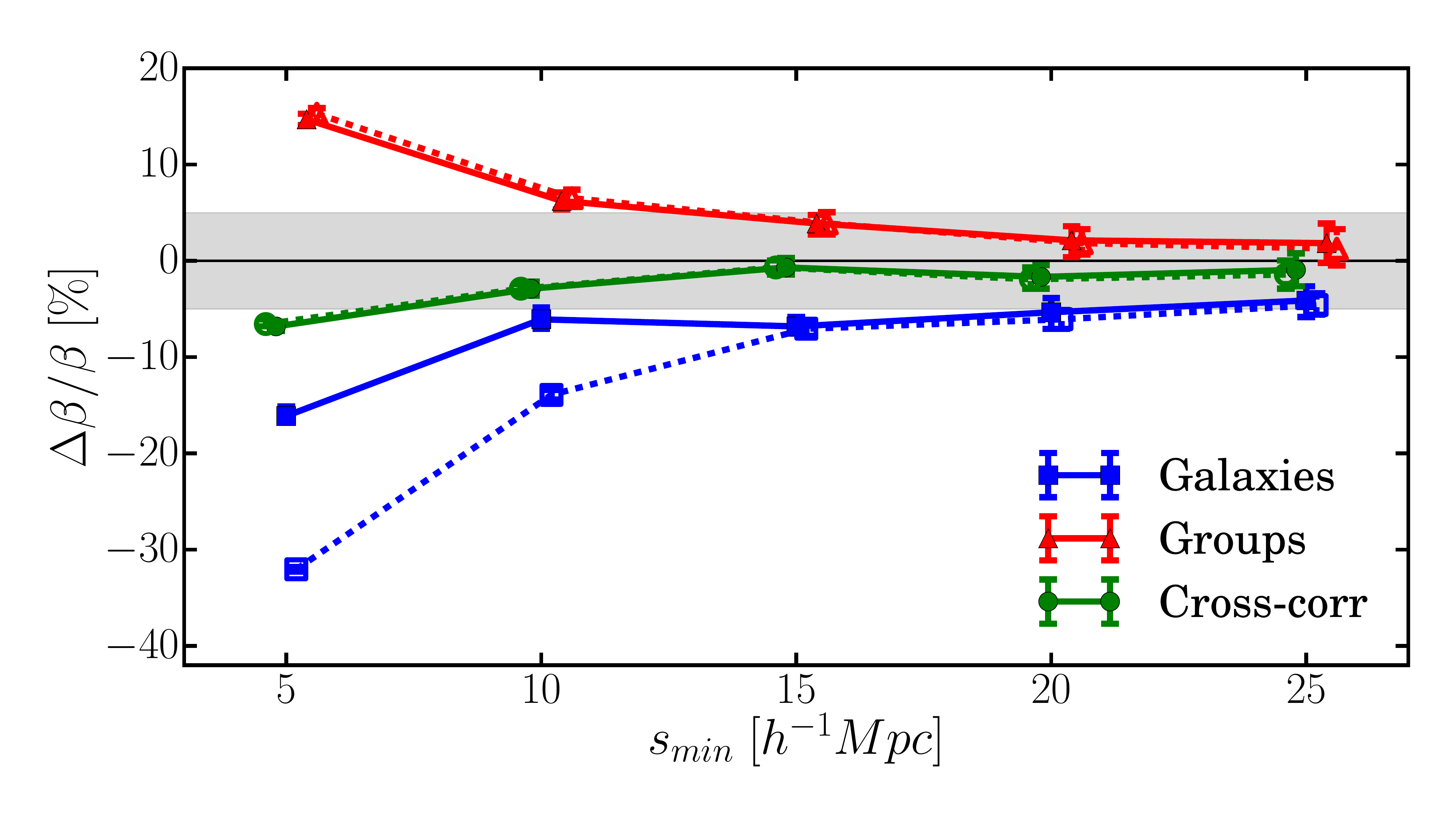}}
			\subfigure{\includegraphics[scale=0.21]{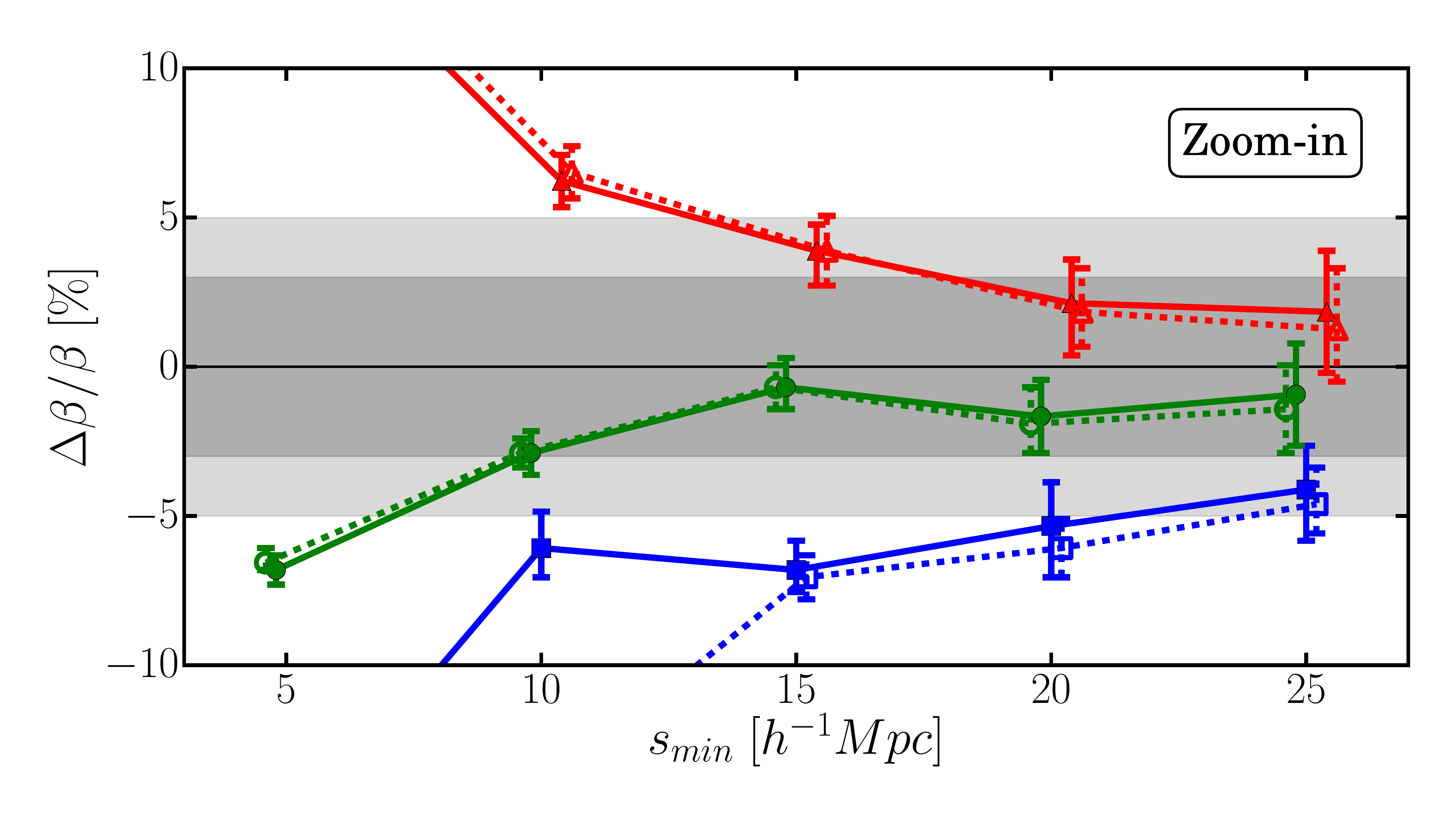}}
			\caption{\small{Same as in Figure
                            \ref{fig:ris_rp-pi} but now fitting
                            truncated monopole and quadrupole moments
                            of the 2PCF in redshift
                            space.}\label{fig:ris_mps-rp}}
\end{figure*}

\subsection{Fits to the standard multipole moments}\label{sec:mps_diag}

We perform here joint fits of the monopole
$\xi^{s,\left(0\right)}\left(s\right)$ and the quadrupole
$\xi^{s,\left(2\right)}\left(s\right)$ of the 2PCF in redshift space. We
do not consider higher order moments, which are too noisy. 
To reduce scale dependence, rather than fitting the multipoles
directly, we consider the quantity \citep[e.g.][]{delatorre13}  
	\begin{equation}
		y^{s,(\ell)}(s)=s^2\,\xi^{s,(\ell)}(s),	\label{eq:s2mps}
	\end{equation}
which we fit between a varying minimum scale $s_{\rm min}$ and a maximum
separation $s_{\rm max}=80\mpcoh$. Also in this case we limit the
likelihood function in equation \eqref{eq:likelihood} to the diagonal
elements of the covariance matrix.  As in the previous case, we
fit using both the full Dispersion model and the simple linear model
only, with results plotted as solid and dashed lines,
respectively, in Figure \ref{fig:ris_mps}.  

The measurements show a different
behaviour, when compared to the fits to the full $\xirp$.  Overall,
the results are much more sensitive to the minimum fitting scale
$s_{\rm min}$.  Also, the linear model gives highly biased results whenever
galaxies are involved (either using their auto-correlation or
the group-galaxy cross-correlation), producing a meaningful result
only when applied to groups alone. 

The measurements from the galaxy auto-correlation (blue
line) and the group-galaxy cross-correlation (green line) are now very
similar, with an overall systematic error on $\beta$, which remains
confined to values smaller than $5\%$. 
On the other hand, the estimates from the group auto-correlation
function have a highly scale-dependent behaviour, with a significant
positive bias when including scales below 10 $\mpcoh$. All
approaches converge to a systematic (negative) error of $ 3-5 \%$
when using only scales $>25\mpcoh$.

We note how in this case, compared to the analysis using the full
$\xi^{s}(r_p,\pi)$,
there is no clear indication that one method performs better than
another.  We interpret this as the consequence of the projection of the 2PCF onto the Legendre
polynomials, which re-distributes over all scales $s$ the non-linear effects
originally mostly confined to small transverse scales
$r_p$. As a consequence, limiting the fitting range to be
above a given $s_{\rm min}$ does not eliminate such small-scale contribution.

\subsection{Fits to the truncated multipole moments}\label{sec:ris_mps-rp_diag}

As described in Section \ref{sec:dispersion}, we apply now the 
truncated multipole moments (equation \ref{eq:projmps_rpcut}).
These have been specifically defined as to eliminate the contribution
of small transverse scales $r_p$, which, as we have just seen, affect all separations $s$. 
As in the previous section, we perform a joint fit of the quantity in equation
\eqref{eq:s2mps} for the case of the truncated monopole $\ell=0$ and quadrupole
$\ell=2$ using only diagonal errors and for the usual varying ranges in scale. 

The results, plotted in Figure \ref{fig:ris_mps-rp},  show a quite
different situation from that of the two previous sections.  
The estimates using the galaxy auto-correlation function (blue
lines) are not improved compared to using the full
multipoles, but maintain a typical negative systematic error of $\sim
5\%$.  Conversely, the estimates using the group auto-correlation (red lines)
show a positive bias for any fitting range, converging to percent errors 
when including only scales larger than 15 $\mpcoh$.  Finally, in
this case the fits to the cross-correlation are surprisingly stable,
almost independently of the fitted range, with an expected systematic
error of 3 \% or less for $s_{\rm min}>10\mpcoh$.  

As evident from the dashed lines, using the linear model alone gives
in general virtually identical results to the fit using the dispersion
model. The only marginal exception is the case of the galaxy
auto-correlation.  This overall behaviour indicates how the truncated
multipoles are able to suppress the weight of small-scale
non-linearities, thus making the role of the dispersion factor
negligible.  In the case of the galaxy auto-correlation, there is
still a difference between the two approaches when including scales
below $15\mpcoh$ in the fit, indicating the stronger effect of
non-linearities in this case, compared to when groups are involved.
Overall, we can conclude that the newly defined statistic of truncated
multipole moments is helpful in reducing the impact of non-linearities
on all scales in the measurements.

\subsection{Full covariance matrix analysis}	\label{sec:results_cov}																

\begin{figure*}
		\centering
			\subfigure{\includegraphics[scale=0.21]{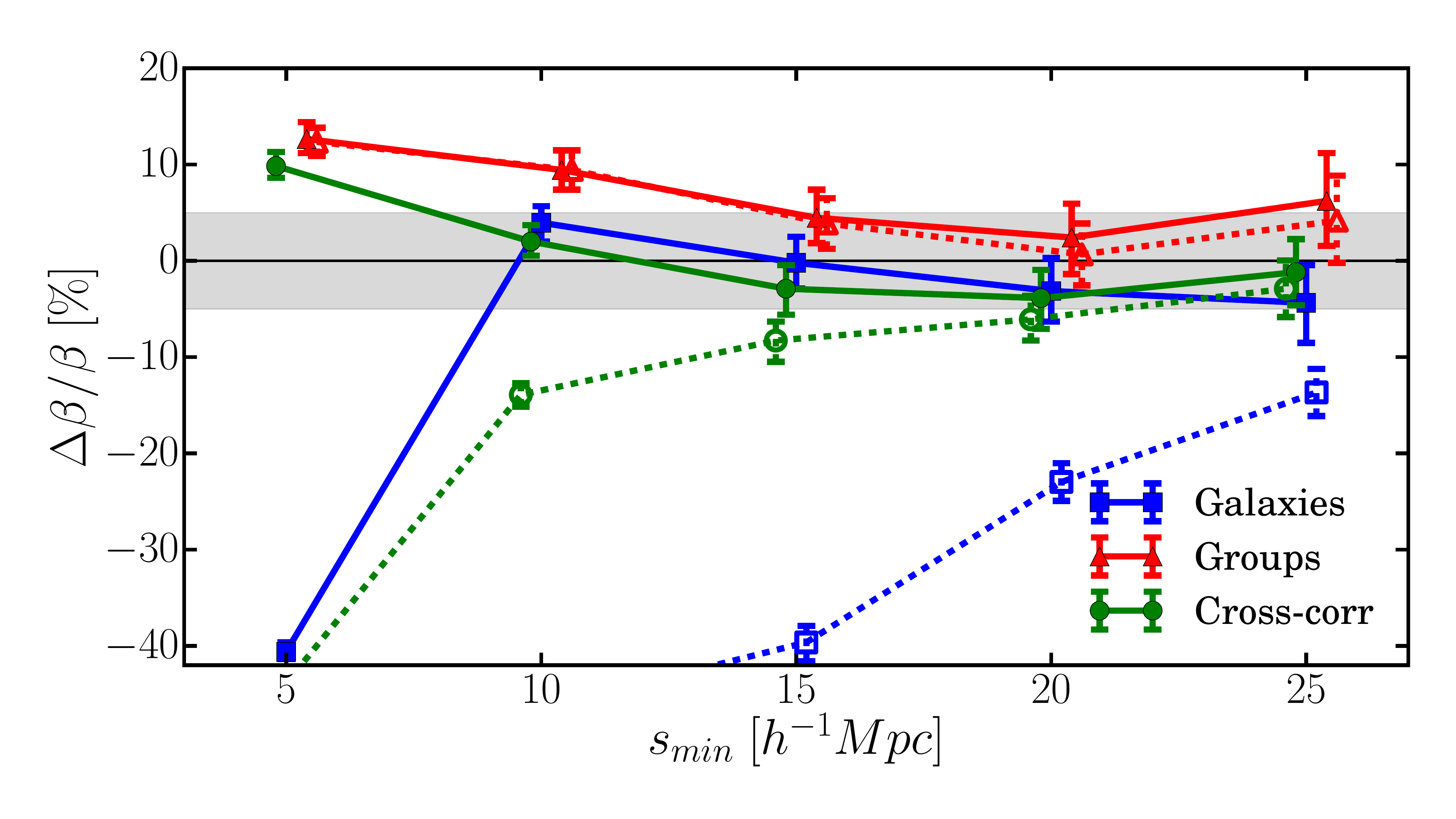}}
			\subfigure{\includegraphics[scale=0.21]{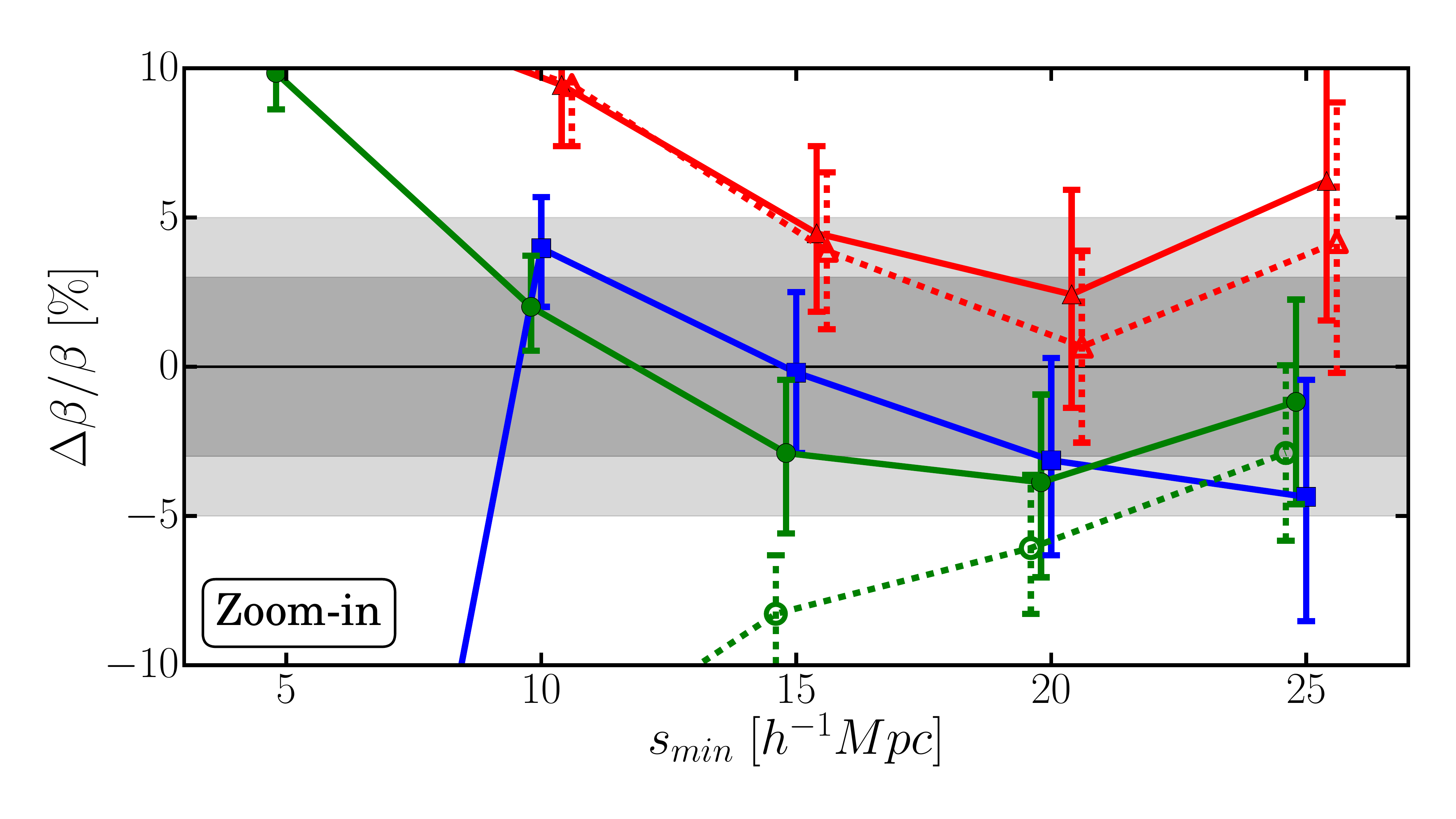}}
			\caption{\small{Same as in Figure
                            \ref{fig:ris_mps} but now fits are
                            performed using the joint covariance
                            matrix of standard monopole and quadrupole
                            moments of the 2PCF in redshift
                            space.}\label{fig:mps_cov}}
\end{figure*}

\begin{figure*}
		\centering
			\subfigure{\includegraphics[scale=0.265]{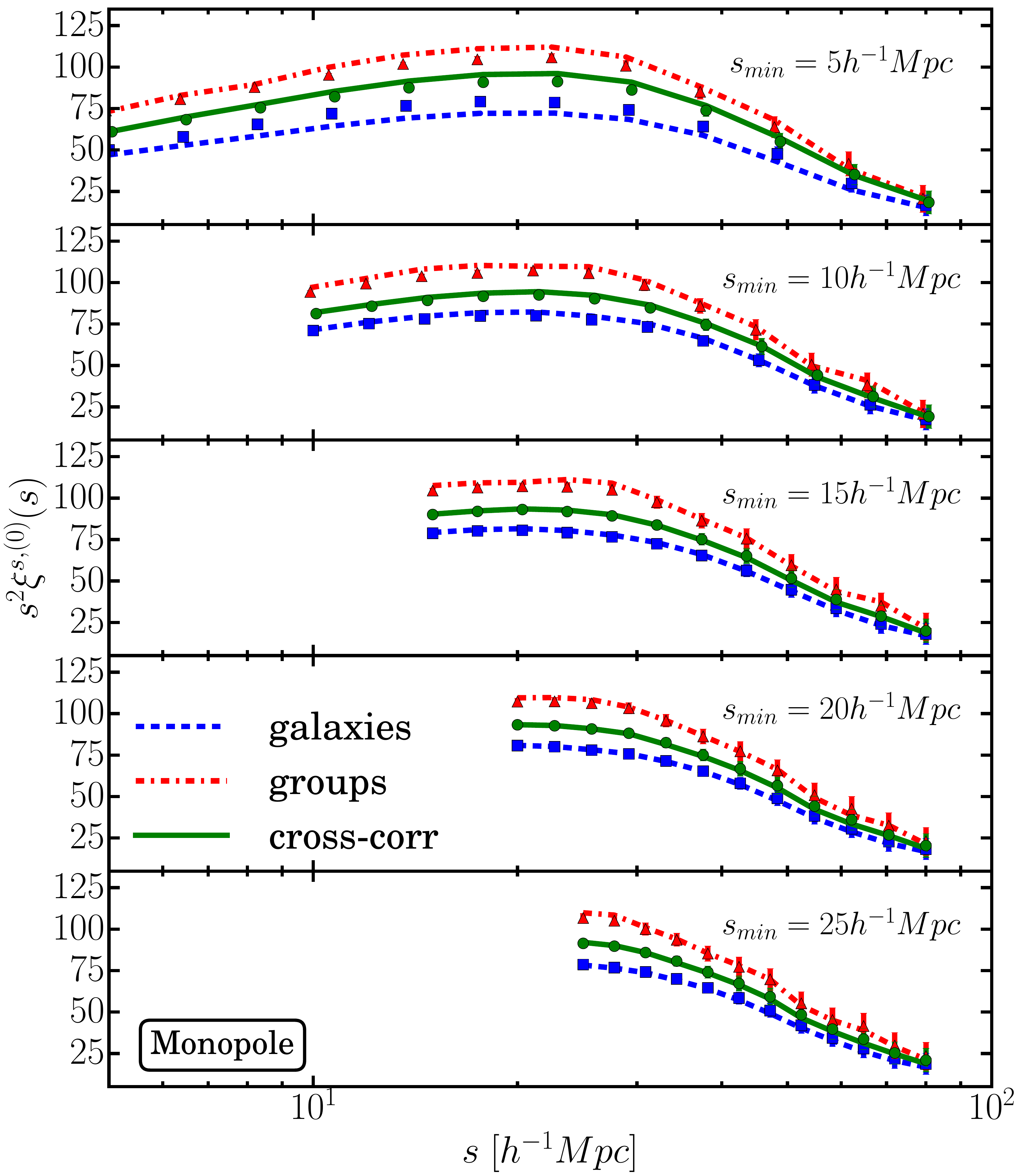}}
			\subfigure{\includegraphics[scale=0.265]{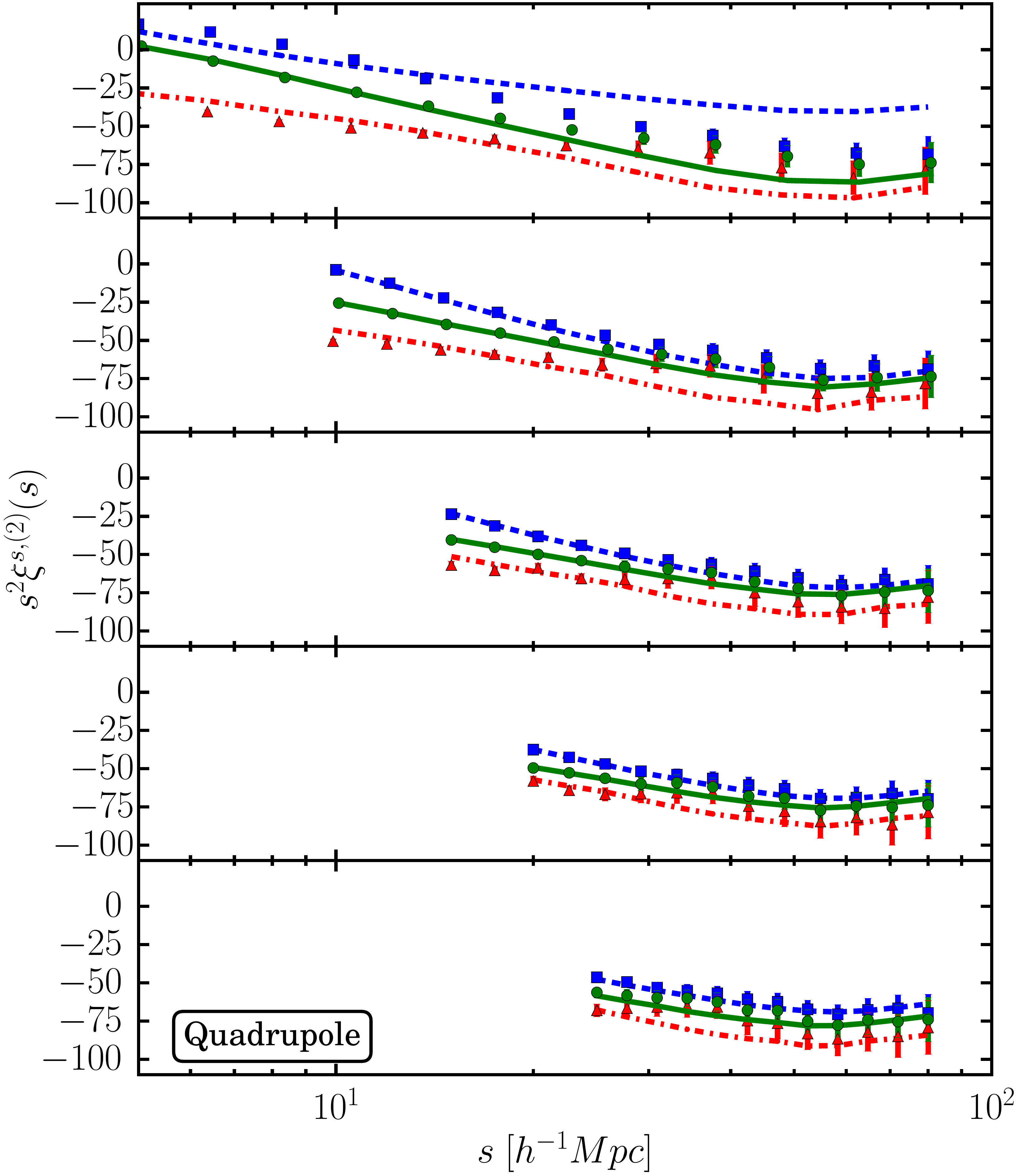}}
			\caption{\small{Measurements (points with error-bars) and the related best fit models (lines) for the standard monopole (left panels) and quadrupole (right panels). Blue squares and dashed lines represent the auto-correlation of (HOD, $L>1L*$) galaxies, red triangles and dash-dotted lines represent the auto-correlation of ($M>10^{13}h^{-1}M_\odot$ dark matter haloes) groups while their cross-correlation is shown through green circles and continuous lines. Each row shows fit performed using the Dispersion model and full covariance matrix at different minimum fitting scales $s_{min}$.}\label{fig:best-fit_cov_mps_ac_gal}}
\end{figure*}

\begin{figure*}
		\centering
			\subfigure{\includegraphics[scale=0.21]{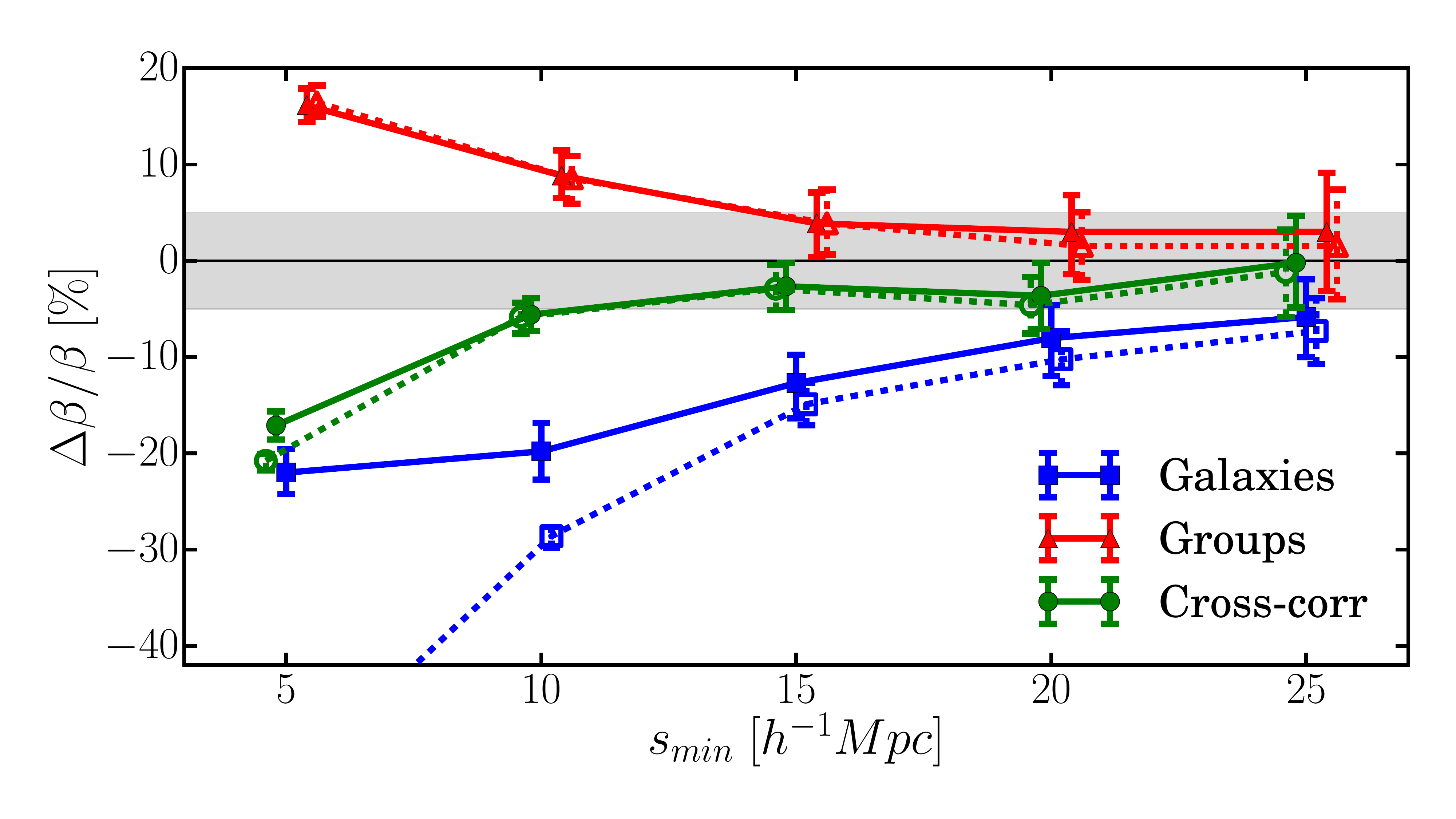}}
			\subfigure{\includegraphics[scale=0.21]{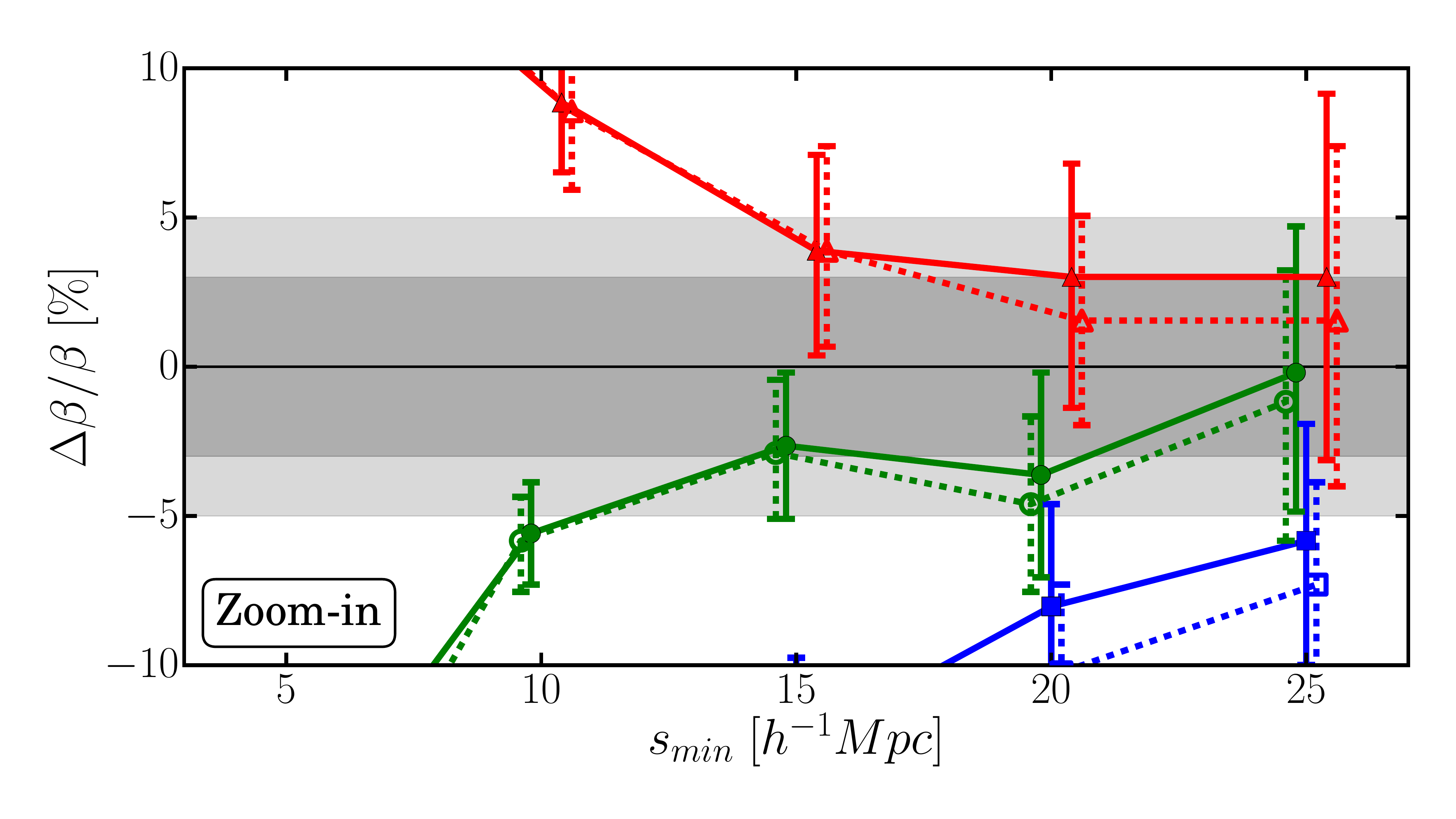}}
			\caption{\small{Same as in Figure
                            \ref{fig:ris_mps-rp} but now fits are
                            performed using the joint covariance
                            matrix of truncated monopole and
                            quadrupole moments of the PCF in redshift
                            space.}\label{fig:ris_cov_mps-rp}}
\end{figure*}

\begin{figure*}
		\centering
			\subfigure{\includegraphics[scale=0.265]{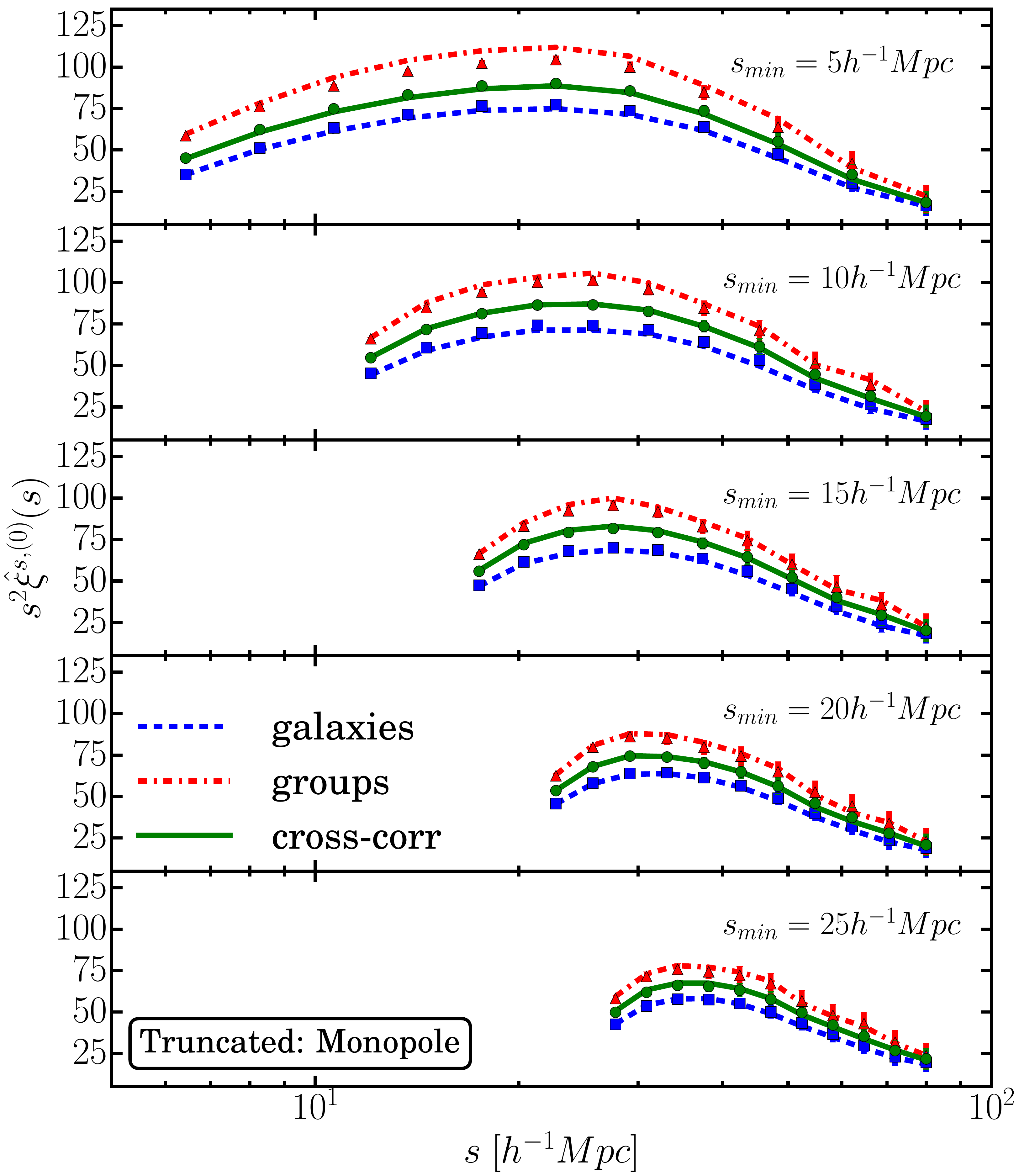}}
			\subfigure{\includegraphics[scale=0.265]{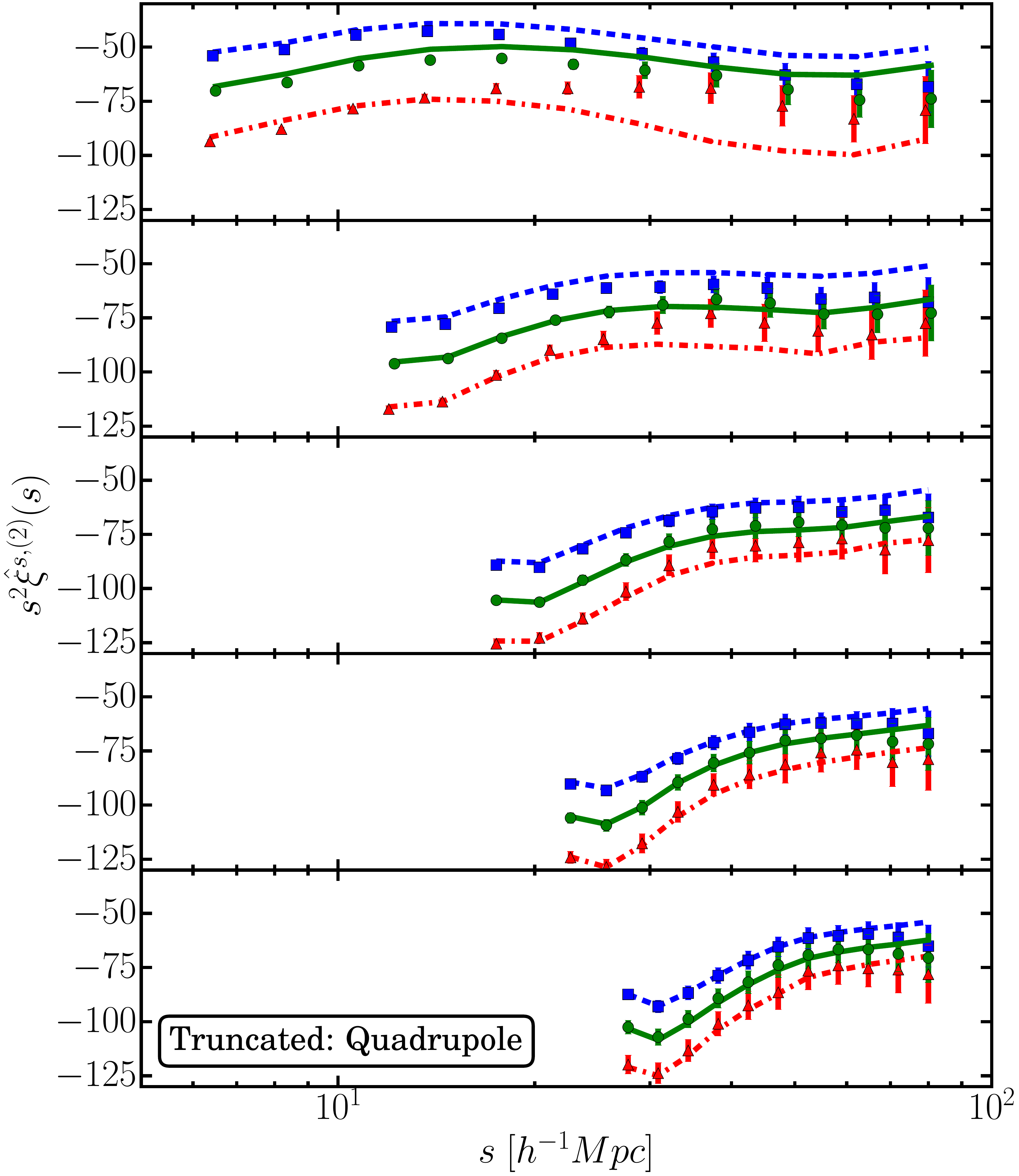}}
			\caption{\small{Same as in Figure
                            \ref{fig:best-fit_cov_mps_ac_gal} but for
                            truncated multipole
                            moments.}\label{fig:best-fit_cov_mps-rp_ac_gal}}
\end{figure*}

Measurements of the two-point correlation function in two different
bins $i$ and $j$ are, in general, correlated with each other.  Keeping
in mind that the estimate and use of a proper covariance matrix is a
non-trivial issue (see e.g. \citejap{delatorre12}), we explore
here the impact of including the full covariance matrix in the
analysis of the standard and truncated multipole
moments. Specifically, we estimate and use the joint covariance matrix
of the truncated monopole $(\ell=0)$ and quadrupole $(\ell=2)$. This
means measuring not only the covariance between the measurements of
the monopole $y^{s,(0)}(s)$ and quadrupole $y^{s,(2)}(s)$ in two
different bins separately but also the cross-covariance between the
measurement of the monopole $y^{s,(0)}(s)$ in a given bin $i'$ and
that of quadrupole $y^{s,(2)}(s)$ in bin $j'$. To do this we store the
measurements of the monopole $y^{s,(0)}$ and the quadrupole
$y^{s,(2)}$, concatenating them into a single array. The covariance
matrix is then measured using the definition in equation
\eqref{eq:cov_sv} from 27 sub-sample realizations.  We recall
  that to avoid a singular covariance matrix and keep a good spatial
  resolution in our measurements we keep fixed the number of bins in
  the fitting range $[s_{min},80]\mpcoh$, independently of $s_{min}$.

We repeat the analysis of standard and truncated multipole
moments of the 2PCF but now including the full data covariance matrix in
the fitting procedure. The best-fitting models are presented in Figures
\ref{fig:best-fit_cov_mps_ac_gal} and
\ref{fig:best-fit_cov_mps-rp_ac_gal}, and the systematic errors on
$\beta$ in Figures \ref{fig:mps_cov} and \ref{fig:ris_cov_mps-rp}. In
general, we find the behaviours with minimum fitting scale for the
different types of correlation to be similar to the case where only
diagonal elements of the covariance matrix are used. The statistical
errors on $\beta$ are however larger, as expected, and the detailed
dependence on minimum fitting scale slightly noisier, consistently with
the increased statistical errors. The only marginal difference in
systematics compared to the diagonal covariance case, is in the galaxy
auto-correlation where we note that both Kaiser and Dispersion models
recover lower values of $\beta$ than previously, in particular for
$s_{min}\le20\mpcoh$.
This analysis confirm the results obtained previously and based on using
diagonal covariance matrix. It confirms in particular the significantly
improved performances of the truncated multipole moments of the
galaxy-group cross-correlation with linear Kaiser model.

\subsection{Dependence of the results on the mass threshold of the
  group catalogue}\label{sec:halo-mass}

 \begin{figure*}
		\centering
			\subfigure{\includegraphics[scale=0.19]{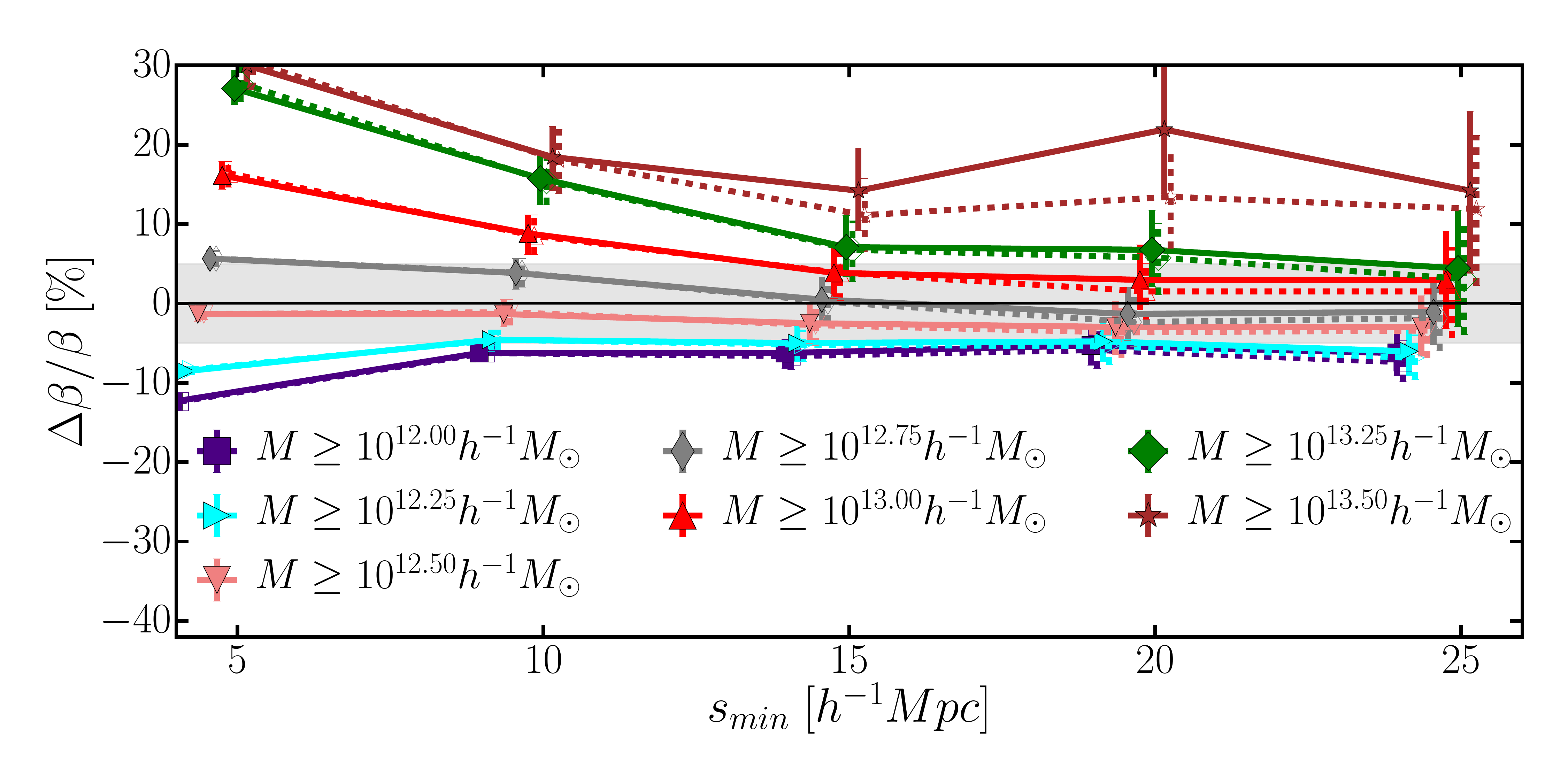}}
			\subfigure{\includegraphics[scale=0.19]{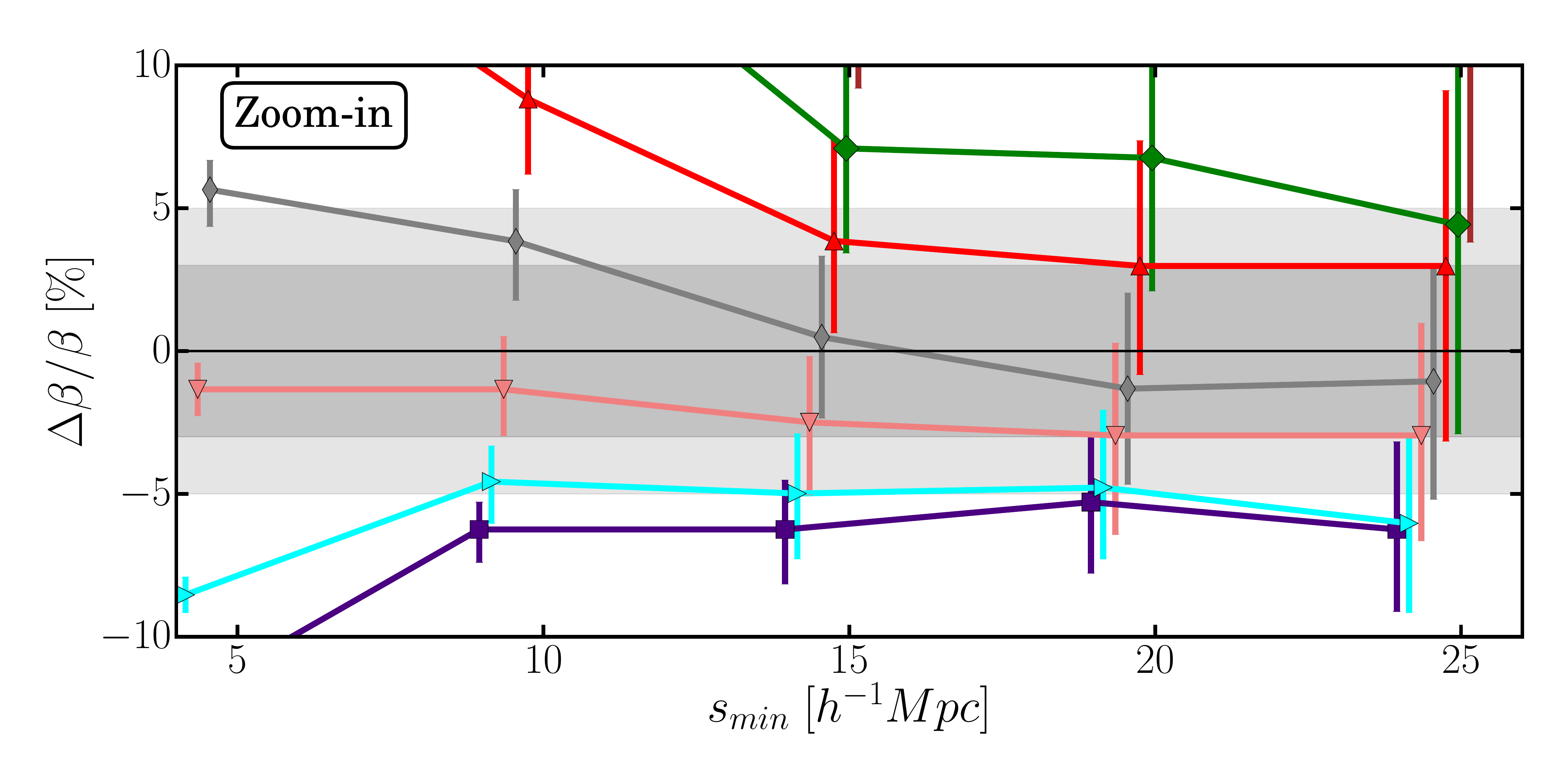}}
			\caption{\small{Global dependence of the
                              systematic errors on $\beta$ obtained
                              from the group 2PCF, when the mass
                              threshold of the group catalogue is
                              increased (Table \ref{tab:group_cats}).
                              These tests use the
                              truncated multipoles and include the
                              full covariance matrix. As in previous
                              plots, the dashed lines and empty
                              markers in the left plot correspond to
                              using the simple Kaiser model fit.  The
                              right panel is a zoom in, including only
                              the Dispersion model curves, to
                              avoid confusion. The red curve ($M>10^{13} h^{-1}M_\odot$)
                              corresponds to the group catalogue used
                              so far in the analyses.}\label{fig:mps-rp_gr_mass}} 
\end{figure*}
\begin{figure*}
		\centering
			\subfigure{\includegraphics[scale=0.19]{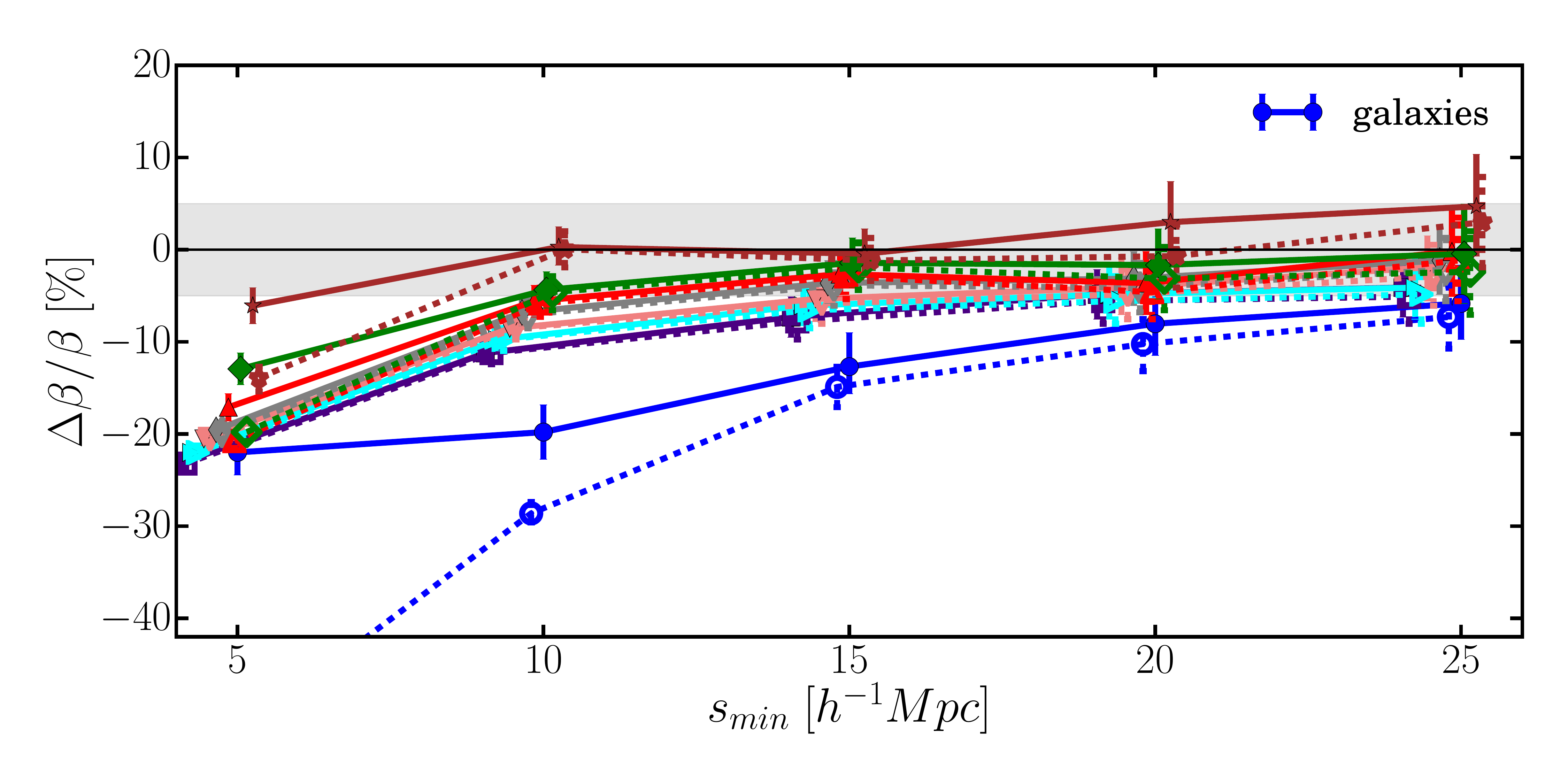}}
			\subfigure{\includegraphics[scale=0.19]{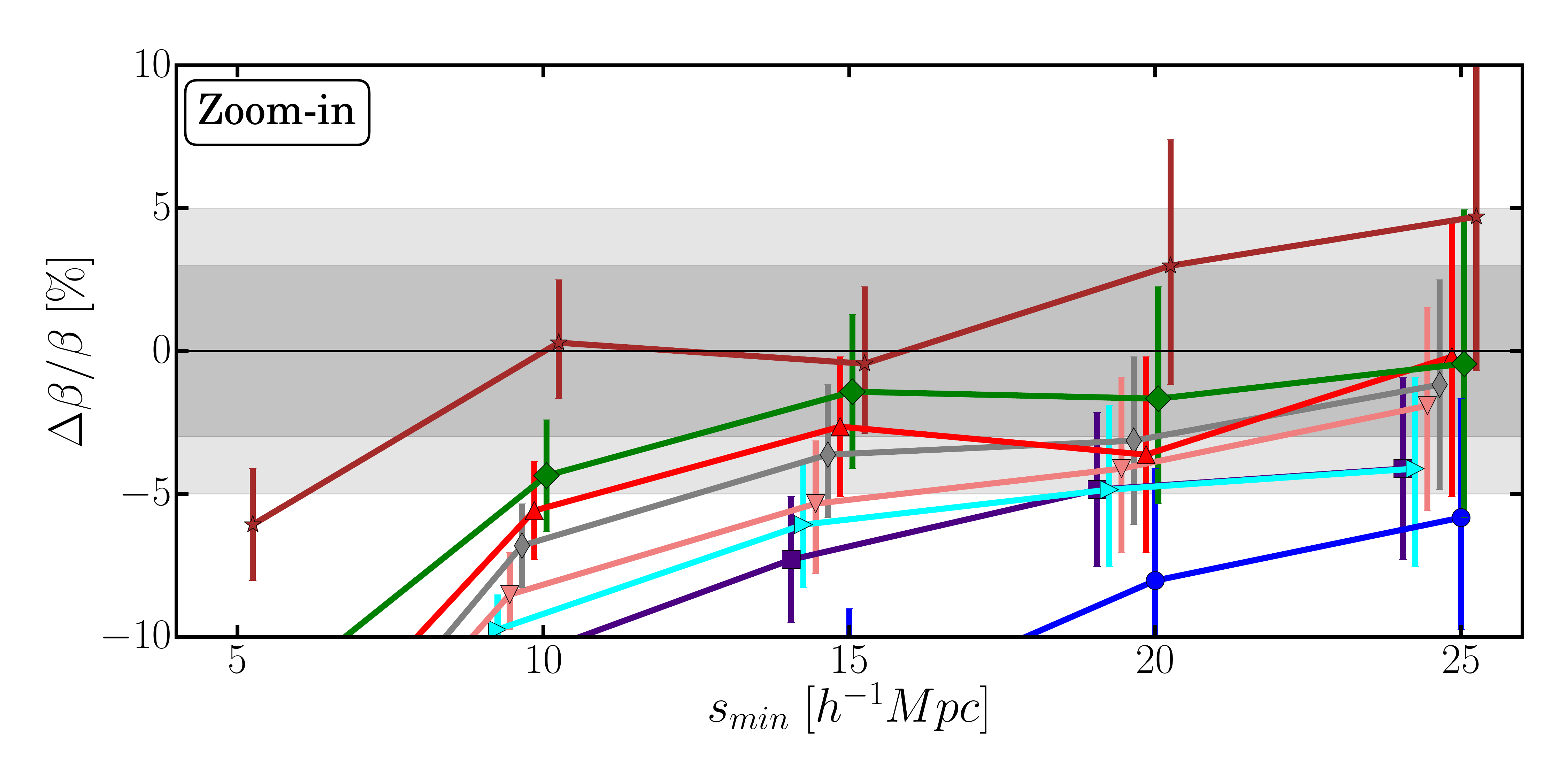}}
			\caption{\small{Same as in Figure
                         \ref{fig:mps-rp_gr_mass}, but now for the
                         cross-correlation of groups and galaxies
                         as defined before.  The values obtained from
                         the galaxy 2PCF are also shown as blue
                         squares and lines, for comparison. The colour coding for other curves is the same as in Figure \ref{fig:mps-rp_gr_mass}.}\label{fig:mps-rp_gal+cross_mass}}
\end{figure*}

\begin{table*}
        			\begin{center}
                			\begin{tabular}{c       c       c	c	c}
                        			\hline
					\hline
				$\mbox{Log}[M_{min}/(h^{-1}M_\odot)]$	&       $N$ 					&       $b_{gr}$					&		$\beta_{gr}^{fid}$		&		$b_{12}$	\\
					\hline	
                   	     				$12.00$				&	$3,439,747$				&	$1.041\pm0.001$			&		$0.5220\pm0.0005$		&		$1.279\pm0.004$	\\
							$12.25$				&	$1,997,835$				&	$1.133\pm0.001$			&		$0.4796\pm0.0004$		&		$1.175\pm0.004$	\\
							$12.50$				&	$1,140,387$				&	$1.254\pm0.002$			&		$0.4333\pm0.0007$		&		$1.061\pm0.004$	\\
							$12.75$				&	$640,598$				&	$1.402\pm0.002$			&		$0.3876\pm0.0006$		&		$0.949\pm0.003$	\\
							$13.00$				&	$350,518$				&	$1.588\pm0.003$			&		$0.3422\pm0.0006$		&		$0.838\pm0.003$	\\
							$13.25$				&	$184,238$				&	$1.810\pm0.004$			&		$0.3002\pm0.0007$		&		$0.735\pm0.003$	\\
							$13.50$				&	$92,403$					&	$2.094\pm0.006$			&		$0.2595\pm0.0007$		&		$0.636\pm0.003$	\\									\hline
					\hline
        				\end{tabular}
      				\caption{\small{Main parameters
                                      for the catalogues of the dark
                                      matter haloes used as proxies
                                      for galaxy groups in Section
                                      \ref{sec:halo-mass}. The first
                                      column contains the minimum mass
                                      limit for each catalogue, $N$ is
                                      the number of objects, $b_{gr}$
                                      is the linear bias,
                                      $\beta_{gr}^{fid}$ is the
                                      fiducial value for the
                                      distortion parameter and
                                      $b_{12}$ is the relative bias
                                      between galaxies and groups as
                                      defined in Equation
                                      \eqref{eq:bias_ratio}.}\label{tab:group_cats}} 
	        		\end{center}
		\end{table*}

So far we have used as ``groups'' a catalogue of dark matter haloes with
  $M>10^{13}h^{-1}M_\odot$.  In this section we test how
  strong is the dependence of the main results obtained so far, on
  this mass threshold.  To this end, we create the set of group
  catalogues listed in table \ref{tab:group_cats}.  We limit our tests
  to using the truncated multipole moments of the 2PCF and obviously
  include the full covariance matrix of the data.  The results for the
  group auto correlation are plotted in Figure \ref{fig:mps-rp_gr_mass}. These show a
  monotonic trend of the systematic error, almost independently of the minimum fitting
  scale $s_{min}$. The ``sweet spot'' for which the error is minimised
  appears to correspond to masses around $10^{12.50} h^{-1}M_\odot$.
  This behaviour agrees with the previous result obtained by
  \citet{bianchi12} using halo catalogues from the BASICC simulation.

  A similar trend is seen in the group-galaxy cross correlation (Figure
  \ref{fig:mps-rp_gal+cross_mass}), but in this case there is a 
  dependence on the minimum fitting scale and a general tendency to 
  underestimate the value of the distortion parameter for all group
  catalogues.  From this figure, we see that the group catalogue used
  for most of the tests in the paper (i.e. $M>10^{13} h^{-1}M_\odot$
  -- green line) is fairly representative of the general behaviour of
  the cross-correlation function when used to estimate $\beta$.

\section{Discussion and Conclusions}		\label{sec:conclusion}	  

This work has explored two different ways of improving the accuracy of
the measurement of the growth rate of cosmological structure: by using
the cross-correlation of individual galaxies with groups of galaxies
as well as by using a novel estimator of the two-point statistics in
redshift space, the \emph{truncated multipole moments}. The aim is to
reduce the impact of non-linearities arising from small-scale random
peculiar pairwise velocities, with respect to the usual approach of
using the multipole moments of the galaxy auto-correlation.  We have
used a set of simulated catalogues of galaxies to compare the accuracy
with which the anisotropic 2PCF $\xirp$, its multipole moments $\xil$,
and its truncated multipole moments $\xilt$, allow the recovery of the
RSD parameters $\beta$. In this comparison we compared both linear
theory and the Dispersion model for RSD.

We find that fitting the full anisotropic auto-correlation function of
galaxies underestimates the distortion parameter by about $10\%$,
confirming the results of \citet{bianchi12}. The
group-galaxy cross-correlation reduces this bias to a level of about
$7\text{-}8\%$, and the group auto-correlation (for groups with mass
larger than $10^{13} h^{-1}M_\odot$) provides us with even
less biased results, reaching an accuracy of about $5\%$. As one may
have expected, there is almost no difference between using either linear theory or
the Dispersion model when fitting the group auto-correlation function:
the Finger of God effect on group centroids is rather minor.

The analysis of standard multipole moments gives no clear indications of
the best choice of tracer for RSD. While the galaxy auto-correlation and
the group-galaxy cross-correlation lead to similar results, the group
auto-correlation produces highly scale-dependent measurements of the
distortion parameter. Such complications can mostly be explained by the
fact that strong non-linearities on small transverse scales $r_p$ are
integrated over all scales when projecting the 2PCF on Legendre
polynomials, which are not captured by linear or dispersion RSD
models.  This does not happen when fitting the full $\xi(r_p,\pi)$
after excluding scales below a given $r_p$.

This problem is alleviated by the \emph{truncated multipole moments}
that we have introduced. These provide the most
accurate estimates of the distortion parameter among the present tests, improving the accuracy
over the use of $\xirp$ and $\xil$. In that case, the galaxy
auto-correlation underestimates the distortion parameter by about $5\%$,
while the group auto-correlation becomes gradually less biased when
using larger minimum scales to reach a few percent accuracy on scales
greater than $20\mpcoh$. The group-galaxy cross-correlation, when
similarly fitted with the Dispersion model, produces more stable and
also more accurate measurements of $\beta$, reaching the percent level
of accuracy when fitting scales greater than $15\mpcoh$. A comparison with the
results from the linear model shows how the truncated multipole moments
allow reducing the impact of small-scale non-linearities in RSD
measurements, making possible the analysis with the simple linear model.
In fact, for the galaxy auto-correlation, the limit at which the
Dispersion model breaks down is translated to $15\mpcoh$, compared with
$20\mpcoh$ in the case of the anisotropic 2PCF.

We studied the impact of bin-to-bin covariances in the fitting
procedure and found no significant difference in terms of systematics,
confirming our findings based on using only the diagonal covariance
matrix.

Finally, we have directly tested how our general results may depend on
the mass threshold chosen to define the the group catalogue used for
most of the tests ($10^{13} h^{-1}M_\odot$).  The exercise using a set
of further six group catalogues, with minimum mass thresholds ranging from $10^{12}
h^{-1}M_\odot$ to $10^{13.5} h^{-1}M_\odot$, shows that the
``standard'' group catalogue is fairly representative of the general
trend observed in the systematic errors when using the cross
correlation function.  Additionally, it further confirms the
dependence of the errors on the halo mass threshold, evidenced in
\citet{bianchi12}.

Although none of the methods studied here yields a zero bias, we find
the results encouraging.  Small systematics at the few percent level
arise in RSD from other effects to do with the sky sampling
\citep{delatorre13}, and these are already corrected for by
analysis of mock data. The same approach could be taken in order to
incorporate small systematic errors in the theoretical RSD models being
used, although further work will be required in order to demonstrate
that these systematic offsets themselves are consistent independent of
the true cosmology under study.

The other element of this study that could benefit from extension
concerns the group catalogue. The present work is somewhat idealised in
that the group proxies are dark-matter haloes that are found directly in
the simulation using more information than would be available in a real
galaxy survey. The next step is therefore to repeat this analysis using
a full simulation of the construction of an empirical group catalogue by
linking the simulated galaxies in redshift space (following e.g. 
\citejap{robotham11}). At large group masses, uncertainties in
group centroids should not be large, and so the features of the present
work in terms of small Fingers of God should be reproduced.

\section*{Acknowledgements}

We thank Julien Bel for discussions and useful suggestions. LG and FM
acknowledge support of the European Research Council through the
Darklight ERC Advanced Research Grant (\# 291521).  SdlT acknowledges
the support of the OCEVU Labex (ANR-11-LABX-0060) and the A*MIDEX
project (ANR-11-IDEX-0001-02) funded by the 'Investissements d'Avenir'
French government program managed by the ANR.

\setlength{\bibhang}{2.0em}
\setlength\labelwidth{0.0em}


\end{document}